\documentclass[prl,a4paper,twocolumn,nofootinbib,superscriptaddress,preprintnumbers,floatfix]{revtex4-1}

\usepackage[utf8]{inputenc}
\usepackage{amssymb}
\usepackage{amsmath}
\usepackage{graphicx}
\usepackage{psfrag}
\usepackage{latexsym}
\usepackage{hyperref}

\newcommand{\bea}{\begin{eqnarray}}
\newcommand{\beal}[1]{\begin{eqnarray}\label{#1}}
\newcommand{\eea}{\end{eqnarray}}
\newcommand{\be}{\begin{equation}}
\newcommand{\bel}[1]{\begin{equation}\label{#1}}
\newcommand{\ee}{\end{equation}}

\newcommand{\bit}{\begin{itemize}}
\newcommand{\eit}{\end{itemize}}
\newcommand{\ben}{\begin{enumerate}}
\newcommand{\een}{\end{enumerate}}

\newcommand{\eps}{\varepsilon}

\newcommand{\ed}{{\cal E}}

\newcommand{\trel}{{\tau_{\mathrm{rel}}}}

\newcommand{\pa}{{\cal A}}
  
\newcommand{\intd}{{\mathrm{d}}}

\begin{document}

\title{How does relativistic kinetic theory remember about initial conditions?}

\author{Michal P. Heller}
\email{michal.p.heller@aei.mpg.de}
\affiliation{Max Planck Institute for Gravitational Physics, Potsdam-Golm, D-14476, Germany}
\affiliation{National Centre for Nuclear Research,
  00-681 Warsaw, Poland}

\author{Viktor Svensson}
\email{viktor.svensson@aei.mpg.de}
\affiliation{National Centre for Nuclear Research,
  00-681 Warsaw, Poland}
\affiliation{Max Planck Institute for Gravitational Physics, Potsdam-Golm, D-14476, Germany}

\begin{abstract}

Understanding hydrodynamization in microscopic models of heavy-ion collisions has been an important topic in current research. Many lessons obtained within the strongly-coupled (holographic) models originate from the properties of transient excitations of equilibrium encapsulated by short-lived quasinormal modes of black holes. This paper aims to develop similar intuition for expanding plasma systems described by a simple model from the weakly-coupled domain, the Boltzmann equation in the relaxation time approximation. We show that in this kinetic theory setup there are infinitely many transient modes carrying information about the initial distribution function. They all have the same exponential damping set by the relaxation time but are distinguished by different power-law suppressions and different frequencies of oscillations, logarithmic in proper time. Finally, we also analyze the resurgent interplay between the hydrodynamics and transients in this setup.

\end{abstract}

\maketitle

\noindent \emph{1. Introduction.--} The success of hydrodynamics as a part of the phenomenological description of data obtained in ultrarelativistic heavy-ion collision experiments at RHIC and LHC has triggered significant theoretical interest in understanding the transition to hydrodynamic regime -- hydrodynamization -- from a microscopic standpoint~\cite{Heinz:2004pj,CasalderreySolana:2011us,Busza:2018rrf}. There are two kind of setups in which this outstanding problem has been addressed to date: strongly-coupled models based on holography (also known as AdS/CFT correspondence or gauge-gravity duality)~\cite{Maldacena:1997re,Witten:1998qj,Gubser:1998bc} and weakly-coupled setups based on kinetic theory (Boltzmann equation),~see,~e.g.,~Refs.~\cite{Florkowski:2017olj,Romatschke:2017vte} for a review of some of these developments. The aim of the present paper is to apply intuitions developed using holographic methods to expanding plasma setups described within kinetic theory and in this way compare the two microscopic mechanisms for hydrodynamization.

We will be studying longitudinally expanding plasma systems in $1+3$ dimensions with the assumption of boost-invariance along the expansion axis $z$, see Ref.~\cite{Bjorken:1982qr}, and conformal equation of state relating matter energy density $\cal E$ and its equilibrium pressure $\cal P$ as ${\cal E} = 3 \, {\cal P}$. Assuming local thermalization at late proper time $\tau = \sqrt{t^{2} - z^2}$, the energy-momentum tensor of matter is fully characterized by one dimensionful number $\Lambda$ setting the prefactor in the asymptotic scaling of energy density with proper~time~\cite{Chesler:2009cy}
\begin{equation}
\label{eq.epsBjorken}
{\cal E} \Big|_{\tau \rightarrow \infty} = \frac{\Lambda^4}{\left( \Lambda \, \tau \right)^{4/3}}\,.
\end{equation}
As reviewed in Sec.~3, power law corrections to the above equation have an interpretation as a hydrodynamic gradient expansion and, at least superficially, do not require new information about initial conditions. This raises the puzzle encapsulated by the title of our paper. The microscopic dynamics in the setup of interest is captured by the distribution function $f(x,p)$, which is a non-negative function of spacetime position $x^{\mu}$ (in the present setup only $\tau$ will matter) and the on-shell particle 4-momentum~$p^{\mu}$ (we are assuming here for simplicity massless microscopic constituents). The energy-momentum tensor of the underlying matter is given by the second moment of the distribution function
\begin{equation}
\label{eq.Tmunu}
T^{\mu \nu} = \int \mathrm{d}P \, p^{\mu} \, p^{\nu} \, f(x,p), 
\end{equation}
where $\mathrm{d}P$ stands for the phase space measure defined in Eq.~\eqref{eq.dP}. Local energy density ${\cal E}(\tau)$ in Eq.~\eqref{eq.epsBjorken} is simply equal to $- T^{\tau}_{\, \, \tau}$. The distribution function itself solves a first order partial differential equation (the Boltzmann equation) of the form
\begin{equation}
\label{eq.Boltzmann}
p^{\mu} \partial_{\mu} f(x,p) = {\cal C}[f],
\end{equation}
where the collisional kernel $\cal C$ depends only on the distribution function $f$ at a given spacetime point $x^{\mu}$. As a result, in order to solve the initial value problem one needs to know the distribution on some time-like hypersurface (here taken to be a hypersurface of constant $\tau$) as a function of 4-momentum~$p^{\mu}$. Such an initial condition contains infinitely many data (dimensionful parameters), which is in stark contrast with the late time behaviour captured by Eq.~\eqref{eq.epsBjorken}. Even with a simplifying assumption of rotational invariance in the transversal plane, the initial distribution function is an arbitrary non-negative function of two variables. To rephrase our title: what kind of corrections to Eq.~\eqref{eq.epsBjorken} carry the vast majority of information about initial conditions set by the initial distribution function?

As already anticipated, we will be interested in answering this question using intuitions developed in the holographic studies of heavy-ion collisions. In fact, holography shares one key feature with the present setup: the microscopic description is naturally formulated using variables living in more dimensions than observables (there: correlation functions of operators, here: moments of the distribution function). In holography, if one neglects nonlinear effects, one finds that Eq.~\eqref{eq.epsBjorken} is supplemented by a discrete set of infinitely many corrections of the form
\begin{equation}
\label{eq.QNMs}
\hspace{-3 pt}\delta {\cal E}_{j} \Big|_{\tau \rightarrow \infty} \hspace{-3 pt}= b_{j} \, \tau^{\alpha_{j}} \, e^{-\gamma_{j} \, (\Lambda \, \tau)^{2/3}} \cos{\left( \omega_{j} (\Lambda\, \tau)^{2/3} + \phi_{j} \right)},
\end{equation}
which encapsulate transient excitations undergoing exponential decay with oscillations~\cite{Janik:2006gp,Heller:2014wfa}. Then, at least superficially, an infinite set of amplitudes $b_{j}$ and phases $\phi_{j}$ offers a possibility of encoding information about initial conditions set in the higher dimensional gravitational description. Furthermore, the decay rates $\gamma_{j}$ and oscillation frequencies $\omega_{j}$ are related to positions of single-pole singularities in complexified frequency and at zero momentum in the Fourier-transformed retarded two-point function of the energy-momentum tensor in global thermal equilibrium~\cite{Janik:2006gp}. The character of these singularities being single poles has been understood as the hallmark feature of strongly-coupled setups, see, e.g., Ref.~\cite{Hartnoll:2016apf}. On the gravity side, these singularities are the aforementioned transient quasinormal modes of dual black holes~\cite{Kovtun:2005ev}.

Preliminary results from Ref.~\cite{Romatschke:2015gic} (see also recent Ref.~\cite{Kurkela:2017xis}) confirm general expectations that the relevant singularities of the energy-momentum tensor in kinetic theory are of branch-cut type, whereas Eq.~\eqref{eq.QNMs} holds for simple poles. Our paper, therefore, is all about understanding how Eq.~\eqref{eq.QNMs} gets modified in the simplest kinetic theory model, considered for example by Ref.~\cite{Romatschke:2015gic}. The only ``microscopic'' parameter in the collisional kernel, the relaxation time, is considered to exhibit a general power-law dependence on temperature (but not on quasiparticles momenta, as in Ref.~\cite{Kurkela:2017xis}), see Eq.~\eqref{eq.taurel}. Therefore, our study includes some of the results of Refs.~\cite{Denicol:2016bjh,Heller:2016rtz,Blaizot:2017lht,Blaizot:2017ucy} as special cases.

We believe the issue we are raising and the setup we are using to address it are interesting for a number of reasons. First and foremost on the motivational front: if one were to search for transient phenomena in heavy-ion collisions or other setups, one would naturally search for excitations of a type given by Eq.~\eqref{eq.QNMs} rather than perturbations of global equilibrium. Furthermore, interplay between hydrodynamics and transients has become a topic of significant interest in the past decade. This includes formulating effective theories of hydrodynamics with a view towards better phenomenological description of experimental data~\cite{Baier:2007ix,Florkowski:2010cf,Martinez:2010sc,Denicol:2012cn,Denicol:2012es,Heller:2014wfa,Stephanov:2017ghc}, applications of resurgence techniques to non-equilibrium setups in which transient modes act as analogues of non-perturbative effects and hydrodynamics represents an asymptotic perturbative expansion~\cite{Heller:2013fn,Heller:2015dha,Basar:2015ava,Aniceto:2015mto,Denicol:2016bjh,Heller:2016rtz,Spalinski:2017mel}, as well as viewing hydrodynamics beyond gradient expansion as a set of special attractor solutions~\cite{Heller:2015dha,Romatschke:2017vte,Spalinski:2017mel,Strickland:2017kux,Romatschke:2017acs,Denicol:2017lxn,Behtash:2017wqg,Blaizot:2017ucy}. Our work is also motivated by ongoing efforts to bridge weak- and strong-coupling approaches using holography with higher-derivative corrections~\cite{Grozdanov:2016vgg,Andrade:2016rln,Grozdanov:2016zjj,Grozdanov:2016fkt,Casalderrey-Solana:2017zyh} and extrapolating kinetic theory predictions from weak to realistic/larger couplings~\cite{Keegan:2015avk,Heller:2016rtz}. Last but not least, our studies are also relevant for attempts to use kinetic theory to map early time dynamics in heavy-ion collisions to hydrodynamics~\cite{Keegan:2016cpi}. The interested reader is invited to consult recent review articles~\cite{Florkowski:2017olj,Romatschke:2017ejr,Alqahtani:2017mhy} for an extended discussion of some of these developments.

\vspace{12 pt}

\noindent \emph{2. Kinetic theory models of interest.--} Following earlier studies in Refs.~\cite{Baym:1984np,Florkowski:2013lya,Romatschke:2015gic,Denicol:2016bjh,Heller:2016rtz,Blaizot:2017lht,Denicol:2017lxn,Blaizot:2017ucy} we consider expanding plasma systems governed by kinetic theory with the collisional kernel ${\cal C}[f]$ in the so-called relaxation time approximation (RTA). In our presentation we will follow the conventions of Ref.~\cite{Florkowski:2013lya}. 

The RTA ansatz was introduced originally in Refs.~\cite{Bhatnagar:1954zz,Anderson:1974a} and constitutes perhaps the simplest kinetic theory model with hydrodynamic behavior. Within this ansatz, the collisional kernel is linear in the distribution function and vanishes when the latter takes the equilibrium form~$f_{0}(x, p)$:
\begin{equation}
\label{eq.CRTA}
C[f] = p \cdot U(x) \, \frac{f(x,p) - f_{0}(x, p) }{\tau_{\mathrm{rel}}}.
\end{equation}
This theory contains one adjustable ``microscopic'' variable, the relaxation time $\tau_{\mathrm{rel}}$, and requires specifying the relevant equilibrium distribution function~$f_0(x, p)$. We take the latter to be of the Boltzmann form, i.e.
\begin{equation}
\label{eq.equilibrium}
f_0(x, p) = \frac{1}{\left( 2\pi \right)^{3}} \exp{\left[- \frac{p \cdot U(x)}{T(x)} \right]}.
\end{equation}
Generalizations to Dirac-Fermi and Bose-Einstein distributions are straightforward. In Eq.~\eqref{eq.equilibrium}, and also in Eq.~\eqref{eq.CRTA}, $T(x)$ is the effective temperature, i.e. the temperature of the equilibrium state with the same local energy density ${\cal E}$. In the present case they are related by
\begin{equation}
\label{eq.Tdef}
{\cal E} = \frac{3}{\pi^{2}} \, T^{4}.
\end{equation}
Furthermore, the unit timelike four-vector $U(x)$ is the flow velocity defined by the Landau frame (Landau matching) condition for the energy-momentum tensor given by Eq.~\eqref{eq.Tmunu}:
\begin{equation}
\label{eq.Landau}
T^{\mu}_{\, \, \nu}\,  U^{\nu} = - {\cal E} \, U^{\mu}.
\end{equation}

The last part in specifying the model is defining the relaxation time. We will specialize to models with the relaxation time~$\tau_{\mathrm{rel}}$ exhibiting power-law dependence on the effective temperature
\begin{equation}
\label{eq.taurel}
\tau_{\mathrm{rel}} = \gamma \, T(\tau)^{-\Delta},
\end{equation}
where the overall constant $\gamma$, dimensionful for $\Delta \neq 1$, will be set to unity and can be always restored based on dimensional analysis / physical grounds. Two values of $\Delta$ stand out: $\Delta = 0$ for which the relaxation time is constant and the theory significantly simplifies and $\Delta = 1$ for which the theory is conformally-invariant. Gradient expansions (at large orders) in such RTA models were considered earlier, respectively, in Ref.~\cite{Denicol:2016bjh} and Ref.~\cite{Heller:2016rtz}.

Let us now specialize to the boost-invariant case~\cite{Bjorken:1982qr}. This flow is easiest to study using coordinates proper time ($\tau$) and spacetime rapidity ($y$) defined by
\begin{equation}
\label{eq.tauy}
t = \tau \, \cosh{y} \quad \mathrm{and} \quad z = \tau \, \sinh{y}.
\end{equation}
Under longitudinal boosts, $\tau$ stays invariant and $y$ gets shifted by a constant. In proper time - rapidity coordinates, components of tensors (e.g. $p^{\mu}$, $U^{\mu}$ or $T^{\mu \nu}$) are boost-invariant as long as they do not depend on $y$.

The kinematics of this simple flow dictates that
\begin{equation}
U = \partial_{\tau}
\end{equation}
and that $T$ be a function of $\tau$ only. The symmetries of the problem leads to an energy-momentum tensor $T^{\mu \nu}$ with three different components, $T^{\tau}_{\,\, \tau}$, $T^{y}_{\,\, y}$ and $T^{1}_{\,\, 1} = T^{2}_{\,\, 2}$ defining (minus) local energy density ${\cal E}(\tau)$, longitudinal pressure ${\cal P}_{L}(\tau)$ and transversal pressure ${\cal P}_{T}(\tau)$ respectively. They are further related by tracelessness (note massless particles) and conservation equations of the energy-momentum tensor, implying 
\small
\begin{equation}
\label{eq.PLandPT}
{\cal P}_{L}(\tau) = - {\cal E}(\tau) - \tau \, \dot{\cal E}(\tau) \,\,\, \mathrm{and} \,\,\, {\cal P}_{T}(\tau) = {\cal E}(\tau) + \frac{1}{2} \tau \, \dot{\cal E}(\tau).
\end{equation}
\normalsize
The natural observable, and a measure of deviations from local thermal equilibrium, is the pressure anisotropy normalized to what would be the equilibrium pressure at the same energy density~\cite{Heller:2011ju,Florkowski:2017olj}, i.e.
\begin{equation}
\label{eq.defA}
{\cal A}(\tau) = \frac{{\cal P}_{T}(\tau) - {\cal P}_{L}(\tau)}{{\cal P}(\tau)},
\end{equation}
where ${\cal P}(\tau) ={\cal E}(\tau) / 3$.
Moving on to the microscopic level, one can take the distribution function to be a function of proper time $\tau$, dimensionless combination $\tau \, p^{y} \equiv \hat{p}^{y}$ and the magnitude of the transversal momentum $p_{T}$. In this parametrization, the Boltzmann equation takes a particularly simple form
\begin{equation}
\label{eq.bifBoltzmannRTA}
\partial_{\tau} f(\tau, \hat{p}^{y}, p_{T}) = \frac{f_0(\tau, \hat{p}^{y}, p_{T}) - f(\tau, \hat{p}^{y}, p_{T})}{\tau_{\mathrm{rel}}(\tau)},
\end{equation}
where we remind the reader that the relaxation time in the general case will be time-dependent and
\begin{equation}
\label{eq.feq}
f_0(\tau, \hat{p}^{y}, p_{T}) = \frac{1}{(2\pi)^{3}} \exp{\left(-\frac{\sqrt{\left(\hat{p}^{y}\right)^{2} + \tau^{2} p_{T}^{2}}}{\tau \, T(\tau)}\right)}.
\end{equation}
Lastly, let us remark that the measure factor in the phase space integration $\mathrm{d}P$ reads
\begin{equation}
\label{eq.dP}
\mathrm{d}P = \frac{2\pi\,p_{T}}{\tau \, p^{\tau}} \, \mathrm{d} \hat{p}^{y}\, \mathrm{d}p_{T},
\end{equation}
where
\begin{equation}
p^{\tau} = \frac{1}{\tau} \, \sqrt{(\hat{p}^{y})^{2} + \tau^{2} p_{T}^{2}}
\end{equation}
and, as a result, the energy density takes the form
\begin{eqnarray}
\label{eq.epsfrom2mom}
&&{\cal E}(\tau) = \int \mathrm{d}P \left(p^{\tau}\right)^2 f(\tau, \hat{p}^{y}, p_{T}) = \nonumber \\ && \frac{2\pi}{\tau^{2}}\int_{0}^{\infty} \mathrm{d}p_{T} \int_{-\infty}^{\infty} \mathrm{d}\hat{p}^{y} \, p_{T}\,\sqrt{(\hat{p}^{y})^{2} + \tau^{2} p_{T}^{2}} \, f(\tau, \hat{p}^{y}, p_{T}).\quad\quad,
\end{eqnarray}

Let us now move on to the initial value problem. As anticipated in the introduction, solving Eq.~\eqref{eq.bifBoltzmannRTA} requires knowing $f$ as a function of two variables, $\hat{p}^{y}$ and $p_{T}$, at some initial time $\tau_{0}$. One can see it in two steps. First, one can write a formal integral solution for the distribution function of the form
\begin{eqnarray}
\label{eq.solformal}
f(\tau,\hat{p}^{y}, p_{T}) =&& D(\tau,\tau_{0}) f(\tau_{0},\hat{p}^{y}, p_{T}) + \nonumber \\
&& \int_{\tau_{0}}^{\tau} \frac{d\tau'}{\tau_{\mathrm{rel}}(\tau')} D(\tau,\tau') f_0 (\tau',\hat{p}^{y}, p_{T}), \quad\,
\end{eqnarray}
where
\begin{equation}
\label{eq.defD}
D(\tau_{2},\tau_{1}) = \exp{\left[ - \int_{\tau_{1}}^{\tau_{2}} \frac{\mathrm{d}\tau'}{\tau_{\mathrm{rel}}(\tau')} \right]}.
\end{equation}
Note that the above expression is exponentially suppressed for $\tau_{2} \gg \tau_{1}$. 

In Eq.~\eqref{eq.solformal}, one should bear in mind that the relaxation time can depend on the effective temperature, see Eq.~\eqref{eq.taurel}, and, through Eqs.~\eqref{eq.Tdef} and~\eqref{eq.epsfrom2mom} on the distribution function at a given instance of proper time. This is resolved by taking the second moment of both sides of Eq.~\eqref{eq.solformal} with respect to $p^{\tau}$, as in Eq.~\eqref{eq.epsfrom2mom}, since this leads to an expression depending only on the effective temperature~$T(\tau)$. Indeed, one then obtains the following integral equation \cite{Baym:1984np,Florkowski:2013lya}
\begin{eqnarray}
\label{eq.Baym}
&&{\cal E}(\tau) \, D(\tau,\tau_{0})^{-1} = {\cal E}_{0}(\tau) + \nonumber\\&&  + \frac{1}{2} \, \int_{\tau_{0}}^{\tau} \frac{\mathrm{d} \tau'}{\tau_{\mathrm{rel}}(\tau')} {\cal E}(\tau') \, D(\tau',\tau_{0})^{-1} H\left(\frac{\tau'}{\tau}\right),
\end{eqnarray}
where 
\begin{equation}
\label{eq.defH}
H(q) = q^{2} + \frac{\arctan{\sqrt{\frac{1}{q^{2}}-1}}}{\sqrt{\frac{1}{q^{2}}-1}}
\end{equation}
and
\begin{eqnarray}
&&{\cal E}_{0}(\tau) = \frac{2\pi}{\tau^{2}}\int_{0}^{\infty} \mathrm{d}p_{T} \int_{-\infty}^{\infty} \mathrm{d}\hat{p}^{y} \times \nonumber\\ && \times p_{T}\,\sqrt{(\hat{p}^{y})^{2} + \tau^{2} p_{T}^{2}} \, f(\tau_{0}, \hat{p}^{y}, p_{T}).
\end{eqnarray}
Eq.~\eqref{eq.Baym} will play the central role in the analysis here. Before we move on to describing new results, a few remarks are in order. First, Eq.~\eqref{eq.Baym} is an integral equation, i.e. the effective temperature at a given instance of proper time depends on the whole temperature history till that moment. Second, ${\cal E}_{0}(\tau)$ feeds in information about the initial distribution function into the temperature profile as a function of proper time. The definition of ${\cal E}_{0}(\tau)$ implies the symmetry ${\cal E}_{0}(\tau) = {\cal E}_{0}(-\tau)$ and a late time expansion of the form
\begin{equation}
\label{eq:eps0series}
{\cal E}_{0}(\tau) = \frac{1}{|\tau|}\left(\eps_{1} + \frac{\eps_{3}}{\tau^{2}} + \frac{\eps_{5}}{\tau^{4}} + \ldots \right),
\end{equation}
i.e. with only even powers in the parentheses and, in general, with infinitely many independent coefficients $\epsilon_{j}$. Third, the function $H(q)$ under the integral is evaluated only for $q\in (0, 1]$, but as we showed with our collaborators in Ref.~\cite{Heller:2016rtz}, its analytic properties on the complex $q$-plane are, in fact, important. The coarse features of $H(q)$ make it similar to a simple linear function, i.e. $2\,q$, but, as we will see in Sec.~3 and Sec.~4, its fine details directly translate into the values of hydrodynamic transport coefficients and transient modes. Finally, Eq.~\eqref{eq.Baym} is in general a strongly nonlinear equation for the effective temperature $T(\tau)$ or, equivalently, local energy density ${\cal E}(\tau)$ because of the temperature-dependent relaxation time $\tau_{\mathrm{rel}}(\tau)$. However, for constant relaxation time, i.e. for $\Delta = 0$ in Eq.~\eqref{eq.taurel}, Eq.~\eqref{eq.Baym} becomes a linear equation for ${\cal E}(\tau)$. This significant simplification will allow us in Sec.~5 to see some beautiful resurgent relations between the hydrodynamic and the transient parts of ${\cal E}(\tau)$. Otherwise, Eq.~\eqref{eq.Baym} can be solved numerically, which we do in Sec.~6 using a refinement of the method from Ref.~\cite{Florkowski:2013lya}.

\vspace{12 pt}

\noindent \emph{3. Gradient expansion.--} In the boost-invariant flow, the hydrodynamic gradient expansion (i.e. expansion in the Knudsen number) is a power series in the ratio of the microscopic dissipation scale, here set by $\tau_{\mathrm{rel}}$, and size of the gradient set by the kinematics to be $\frac{1}{\tau}$. We will call this dimensionless ratio $w$:
\begin{equation}
\label{eq.wdef}
w \equiv \frac{\tau}{\tau_{\mathrm{rel}}}.
\end{equation}
If we use our ansatz for the relaxation time, then $w$ reads
\begin{equation}
w = \tau \, T(\tau)^{\Delta}.
\end{equation}
In the conformally-invariant case, $\Delta = 1$, we recover the $w$ variable introduced in Ref.~\cite{Heller:2011ju}, which justifies the name. In particular, when comparing different solutions of Eq.~\eqref{eq.Baym} we will be looking at normalized pressure anisotropy $\cal A$ defined in Eq.~\eqref{eq.defA} as a function of $w$. Of course, one can still treat $w$ as a function of proper time, i.e. $w(\tau)$, as we will often do below.

The energy density can therefore be formally expanded~as
\begin{equation}
\label{eq.Epsgrad1}
{\cal E}(\tau) = \frac{\Lambda^{4}}{\left( \Lambda \, \tau \right)^{4/3}} \left( 1 + \frac{e_{1}}{w(\tau)} + \frac{e_{2}}{w(\tau)^{2}} + \ldots \right)
\end{equation}
with coefficients $e_{j}$ fixed by $\Delta$ and independent of the initial condition. Alternatively, one can represent the energy density in the equivalent late-time expansion as
\small
\begin{equation}
\label{eq.Epsgrad2}
\hspace{-6 pt}{\cal E}(\tau) = \frac{\Lambda^{4}}{\left( \Lambda \, \tau \right)^{4/3}} \hspace{-2 pt} \left(\hspace{-1 pt} 1 + \frac{\tilde{e}_{1}}{(\Lambda \, \tau)^{1-\Delta/3}} + \frac{\tilde{e}_{2}}{(\Lambda \, \tau)^{2-2\Delta/3}}
+ \ldots \right)\hspace{-1pt},
\end{equation}
\normalsize
where comparison with Eq.~\eqref{eq.Epsgrad1} allows one to relate $\tilde{e}_{j}$'s and $e_{j}$'s. One can deduce from Eq.~\eqref{eq.Epsgrad2} that the allowed range of parameter $\Delta$ is
\begin{equation}
\label{eq.allowedDeltas}
\Delta < 3,
\end{equation}
as otherwise the relaxation time at late times gets too large to allow for a depletion of gradients. Such an effect is seen in Refs.~\cite{Denicol:2014tha,Denicol:2014xca}, in RTA kinetic theory undergoing Gubser flow. The rapid expansion in that setup drives the system away from thermal equilibrium. The case $\Delta>3$ is not explored in this paper, but we note that the eremitic expansion of Ref.~\cite{Romatschke:2018wgi} may be more appropriate in that case. 

Expansions in Eqs.~\eqref{eq.Epsgrad1} and~\eqref{eq.Epsgrad2} translate directly into the large-$w$ expansion of the normalized pressure anisotropy ${\cal A}$:
\begin{equation}
\label{eq.Agrad}
{\cal A} = \frac{a_{1}}{w} + \frac{a_{2}}{w^{2}} + \ldots
\end{equation}
Again, it should be noted that the gradient expansion in Eq.~\eqref{eq.Agrad} does not contain any information about an initial state and in Eqs.~\eqref{eq.Epsgrad1} and~\eqref{eq.Epsgrad2} the only information sits in the asymptotic scaling set by~$\Lambda$. Regarding relation to transport coefficients, the term $a_{1}$ is related to the ratio of shear viscosity $\eta$ to entropy density and the term $a_{2}$ is related to a combination of second order transport coefficients $\tau_{\pi}$ and $\lambda_{1}$, see, e.g., Ref.~\cite{Florkowski:2017olj} for details.

As noted in Ref.~\cite{Heller:2016rtz}, the gradient expansion in RTA kinetic theory can be generated using integration by parts of the integral in Eq.~\eqref{eq.Baym}. First, let us observe that
\begin{equation}
D(\tau',\tau_{0})^{-1} = \tau_{\mathrm{rel}}(\tau') \,\frac{\mathrm{d}}{\mathrm{d}\tau'} D(\tau',\tau_{0})^{-1}.
\end{equation}
The appearance of a derivative allows for repeated application of integration by parts in Eq.~\eqref{eq.Baym}. Focusing only on the relevant integral one gets
\begin{eqnarray}
\int_{\tau_{0}}^{\tau} \frac{\mathrm{d} \tau'}{\tau_{\mathrm{rel}}(\tau')} H\left(\frac{\tau'}{\tau}\right) {\cal E}(\tau') D(\tau',\tau_{0})^{-1} = \nonumber \\
\int_{\tau_{0}}^{\tau} \mathrm{d} \tau' H\left(\frac{\tau'}{\tau}\right) {\cal E}(\tau') \frac{\mathrm{d}}{\mathrm{d}\tau'} D(\tau',\tau_{0})^{-1} = \nonumber \\
H(1) \, {\cal E}(\tau) \, D(\tau,\tau_{0})^{-1} - H\left(\frac{\tau_{0}}{\tau}\right) {\cal E}(\tau_{0}) \nonumber \\
- \int_{\tau_{0}}^{\tau} \mathrm{d} \tau' \frac{\mathrm{d}}{\mathrm{d}\tau'} \left[ H\left(\frac{\tau'}{\tau}\right) {\cal E}(\tau') \right] \, D(\tau',\tau_{0})^{-1}.
\end{eqnarray}
Exactly the same logic can be applied to the final integral appearing in the above equation, which leads to an iterative scheme that can be executed indefinitely. Every subsequent integration by parts is going to generate a term proportional to $D(\tau,\tau_{0})^{-1}$ which at late times is exponentially enhanced over the other term. Gathering such dominant terms and neglecting others in the iterated version of Eq.~\eqref{eq.Baym} leads to a differential relation involving derivatives of $H(q)$ at $q = 1$  and derivatives of ${\cal E}(\tau)$ measured in units of relaxation time. As a result one obtains 
\begin{equation}
\label{eq:baymIBP}
\sum_{j = 1}^{\infty} \left(- \tau_{\mathrm{rel}}(\tau') \, \frac{\mathrm{d}}{\mathrm{d} \tau'} \, \right)^{j} H\left( \frac{\tau'}{\tau} \right) {\cal E} (\tau') \Bigg|_{\tau' = \tau} = 0\, ,
\end{equation}
which needs to vanish up to exponentially small corrections (hence the equality in the equation above). Using this expression with the sum truncated at, say, $j = 3$ and $\cal E(\tau)$ given by the gradient expansion allows us to determine, in this case, $e_{1}$ and $e_{2}$ in Eq.~\eqref{eq.Epsgrad1} and, as a result, $a_{1}$ and $a_{2}$ in Eq.~\eqref{eq.Agrad}. The result reads
\begin{equation}\label{eq:Acoeffs}
a_{1} = \frac{8}{5} \quad \mathrm{and} \quad a_{2} = \frac{88}{105}-\frac{8}{15} \, \Delta.
\end{equation}
Iterating this scheme further allows one to get higher order transport. This approach works the best for the constant relaxation time in which case one can get the lowest 1500 coefficients. We did this by first using Eq.~\eqref{eq:baymIBP} to derive a recursive relation for coefficients $e_{j}$ from Eq.~\eqref{eq.Epsgrad1}, which, for $\Delta = 0$, happen to be the same as coefficients $\tilde{e}_{j}$ appearing in Eq.~\eqref{eq.Epsgrad2}, and solving this relation. Unfortunately, the number of terms generated in Eq.~\eqref{eq:baymIBP} gets significantly bigger and the whole approach slower for generic values of $\Delta$. However, in all the cases we checked it was sufficient to demonstrate that the gradient expansion has a vanishing radius of convergence, as expected on general grounds~\cite{Florkowski:2017olj}. For a temperature-dependent relaxation time the method from Refs.~\cite{Heller:2016rtz,Florkowski:2017olj} and, perhaps, also Ref.~\cite{Denicol:2016bjh} are better suited to get a significant number of terms, e.g. 425 terms in the conformal case ($\Delta = 1$) considered in Ref.~\cite{Heller:2016rtz}. 

\begin{figure}
	\centering
	\includegraphics[width=0.98\linewidth]{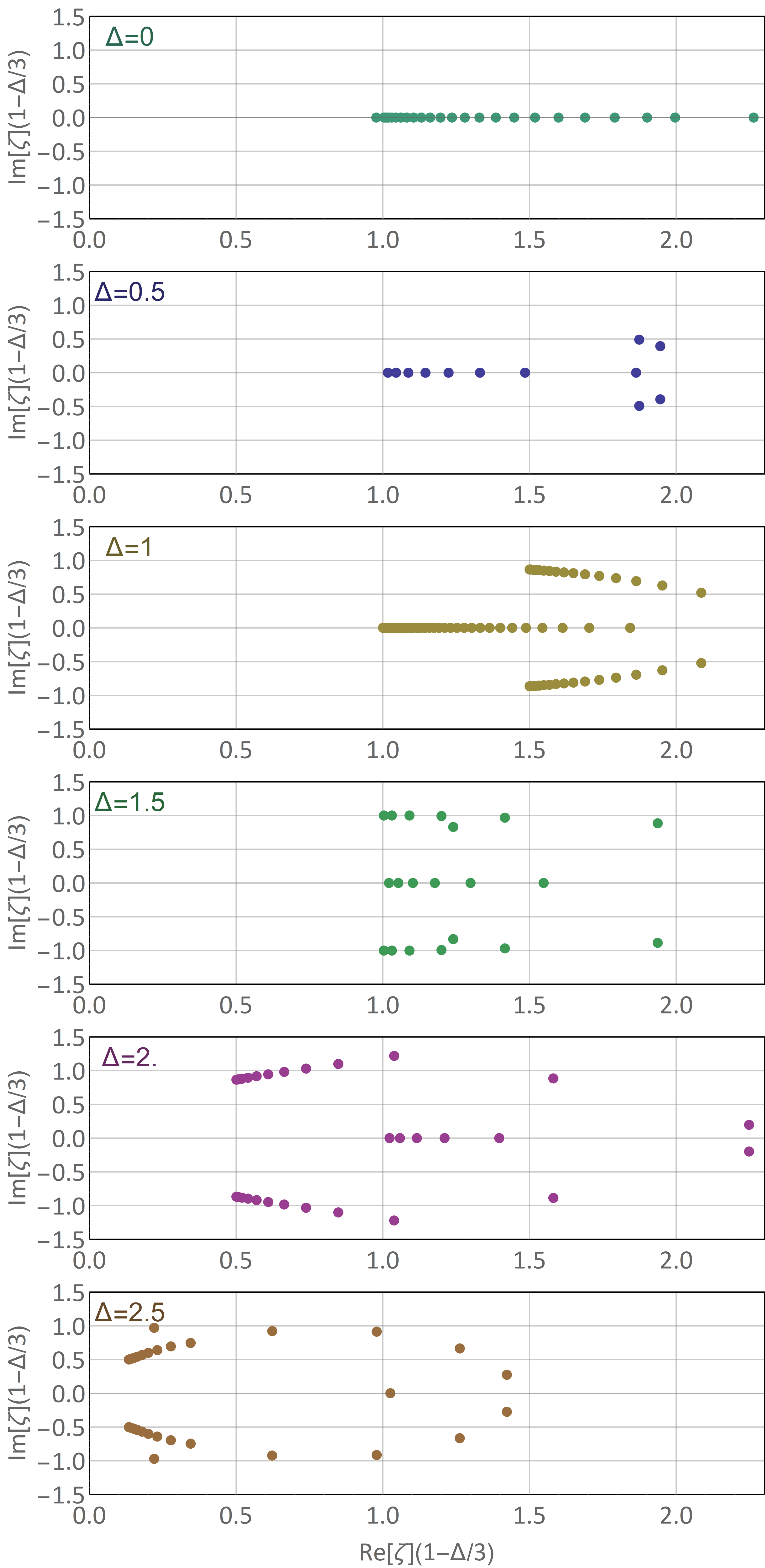}
	\caption{Singularities of the Borel transform of hydrodynamic gradient expansion of $\mathcal{A}$ for sample values of allowed $\Delta$, see Eq.~\eqref{eq.allowedDeltas}. As a method of analytic continuation we use Pad{\'e} approximants. The cases of $\Delta = 0$ and $\Delta = 1$ were studied before in, respectively, Refs.~\cite{Denicol:2016bjh} and~\cite{Heller:2016rtz}. In the plots sequences of poles represent branch cuts, a known feature of Pad{\'e} approximation, see, e.g., Ref.~\cite{PadeCut}. The singularities on the real axis is physical and give rise to transients of the form dictated by Eq.~\eqref{eq:transientAnsatz}. The arguments in Sec.~4 make it clear that this singularity is an infinite set of branch cuts with the same branch point, but of a different order. The singularities off real axis are unphysical and follow from contour deformations in the integral in Eq.~\eqref{eq.Baym}, as explained for $\Delta = 1$ in Ref.~\cite{Heller:2016rtz}. Surprisingly, the unphysical singularities, whose location is at $1+(-1)^{\pm \Delta/3}$, start controlling the radius of convergence of the hydrodynamic series for $\Delta > 2$. }
\label{fig:borelplots}
\end{figure}

A standard way of dealing with asymptotic series is Borel transform, which takes $a_{n}w^{-n}$ to $a_{n} \zeta^n / n!$, and Borel summation which at the level of a series inverts the former operation, see e.g. Ref.~\cite{Florkowski:2017olj}. In Fig.~\ref{fig:borelplots} we show the structure of singularities of the Borel transform of the truncated hydrodynamic gradient expansion for six representative values of $\Delta$. As a way of analytically continuing the Borel transform away from the origin we use the standard symmetric Padé approximation. In Fig.~\ref{fig:borelplots} we always see poles on the real axis and for $\Delta > 0$ also singularities further on the complex plane. As argued in Ref.~\cite{Heller:2016rtz}, the latter are not physical excitations, but rather represent analytic properties of Eq.~\eqref{eq.Baym} with contours of integration over $\tau'$ extended away the real axis. To see this, note that the gradient expansion, computed using Eq.~\eqref{eq:baymIBP}, does not know about the choice of contour between $\tau_0$ and $\tau$ in Eq.~\eqref{eq.Baym}. Different choices of contour will differ by terms coming from singularities of the integrand. By the analytic properties of $H(x)$, such terms will behave as 
\begin{equation}
\delta {\ed} \sim e^{-\frac{1+(-1)^{\pm \Delta/3}}{1-\Delta/3} \, w}.
\end{equation}
While these terms are not to be interpreted as transients, they are nevertheless important. For $\Delta>2$, these will in fact represent the dominant large order behaviour of the gradient expansion. This has similarities with the ghost-instantons studied in quantum mechanics in Ref.~\cite{Basar:2013eka}. Lastly, the movement of the off-axis poles as $\Delta \rightarrow 3$ signals a breakdown of this analysis, as anticipated in the discussion around Eq.~\eqref{eq.allowedDeltas}.

In the present manuscript we will be concerned with singularities lying on the real axis and their relation to transient modes (Sec.~4) and resurgence (Sec.~5).

\vspace{12 pt}

\noindent \emph{4. Transient modes.--}
We have seen that the gradient expansion is universal, independent of initial conditions. In this section, we describe transient corrections to the universal late time behavior. These transient modes come with an overall amplitude and phase that offer the possibility to encode initial information. To this end, we discard $\ed_0$ and set the lower limit of integration $\tau_0/\tau$~to~0. This may seem contradictory, as this removes all initial data. We do this as here we are only concerned with demonstrating how data \textit{can} be stored rather than the particular way a given initial condition \textit{is} stored. We will say more about the matching of initial data to late time modes in Sec.~5. We present here the $\Delta=0$ case as the general case introduces mainly notational, not technical, difficulties.

Since Eq.~\eqref{eq.Baym} contains exponential suppression in the form of $D(\tau,\tau_0)$, a natural ansatz for the energy density is  
\begin{equation} \label{eq:transientAnsatz}
\ed(\tau) =\ed_{ge}(\tau) + \sigma D(\tau,\tau_{0})\ed_{\beta}(\tau),
\end{equation}
where $\ed_{ge}(\tau)$ is the gradient expansion and $\ed_{\beta}(\tau)$ is a power series with leading power $\beta$ (as we will soon see, in general, a complex number), i.e.
\begin{equation}
\label{eq:transpower}
\ed_{\beta}(\tau) = w^{\beta} \left(1+\frac{e_{\beta,1}}{w}+\frac{e_{\beta,2}}{w^2}+\dots\right).
\end{equation}
Inserting this into Eq.~\eqref{eq.Baym} and matching powers of $w$ leads to equations for $\beta$ and $e_{\beta,k}$. In this section, $\beta$ is the object of interest. As described in Sec.~5, for a given $\beta$, the rest of the coefficients $e_{\beta,k}$ are uniquely determined. However, the equations leave $\sigma$ undetermined. Hence, each allowed value of $\beta$ supplies one free parameter where initial data can be stored. Also, it is implicitly assumed in Eq.~\eqref{eq:transientAnsatz} that we sum over all allowed (as we will soon see, infinitely many) values of $\beta$, each with an independent value of $\sigma$.

\begin{figure}[t]
	\centering
	\includegraphics[width=0.9\linewidth]{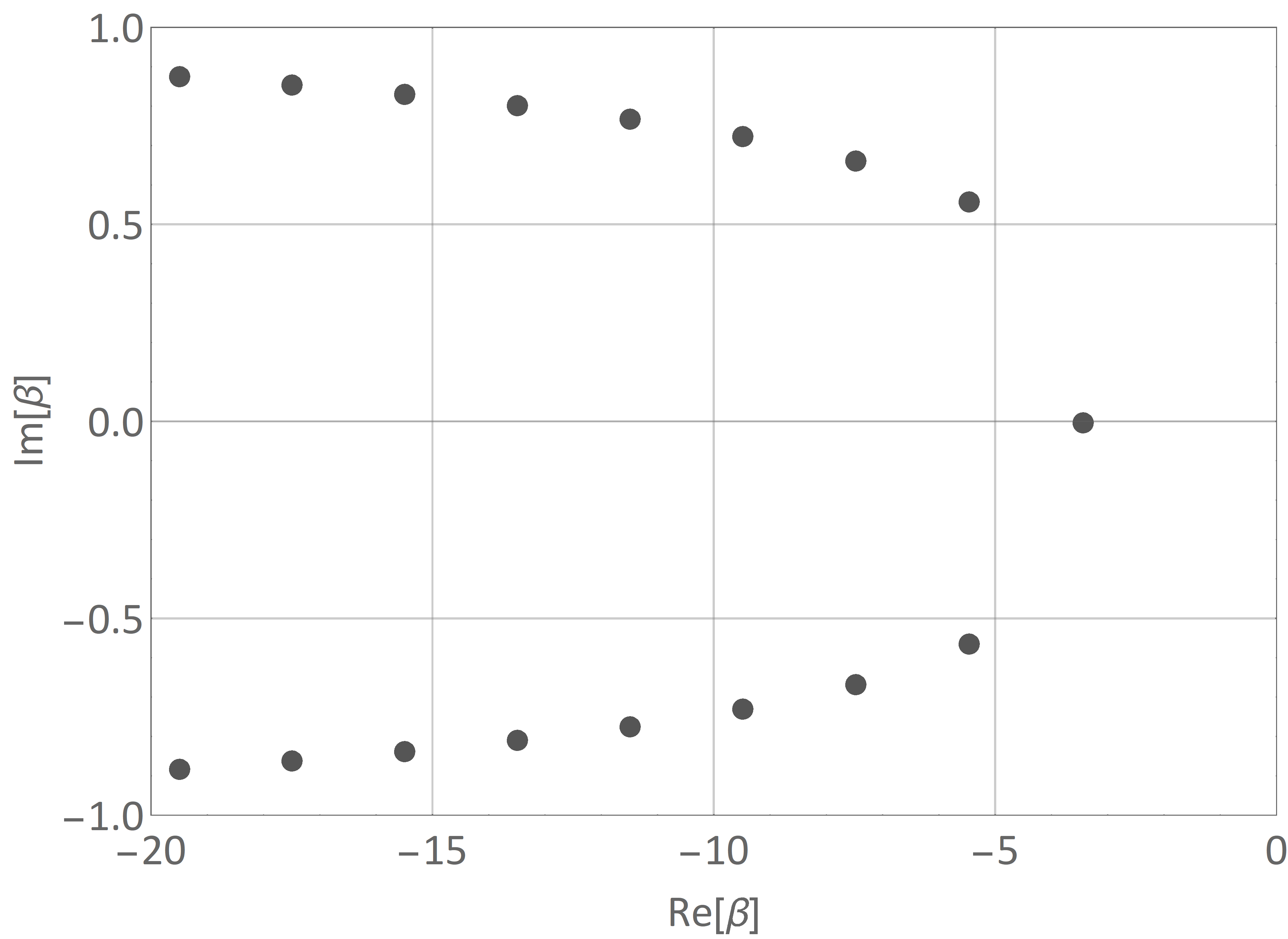}
	\caption{Points in the figure shows roots of the function $M(\beta)$. For $\Delta=0$, each root gives rise to a transient mode of the form $e^{-\tau} \tau^\beta$. For other $\Delta$, the modes behave as in Eq.~\eqref{eq:deltatransient}. The roots with the largest real part will be the dominant ones. The first three are  $\beta_1 \approx -3.4313$, $\beta_{\pm 2} \approx -5.4584 \pm 0.5614 i$,  $\beta_{\pm 3} \approx -7.4746 \pm 0.6648 i$.}
	\label{fig:betas}
\end{figure}

One finds that the $\beta$'s are given by zeros of the function
\begin{equation}
 M(z) \equiv \int_{0}^{1} \mathrm{d}x H\left(x\right) x^{z}.
\end{equation}
Note that the integral converges only for $z>-1$ and that for such $z$, $M(z)>0$. One must analytically continue $M(z)$ to complex $z$ to find any solutions. This can be done by using series expansion for $H$ or the representation
\small
\begin{equation}
\hspace{-7pt}M(z)= \frac{\, _3F_2\left(1,\frac{z}{2}+2,\frac{z}{2}+2;\frac{z}{2}+\frac{5}{2},\frac{z}{2}+3;1\right)}{2 \, z^2+14 \, z+24}+\frac{1}{2 (z+4)}.
\end{equation}
\normalsize
Solutions to $M(z)=0$ are shown in Fig.~\ref{fig:betas}. The string of zeros seems to continue indefinitely, leading us to believe that there are an infinite number of allowed $\beta$'s. One is purely real and the rest come in conjugate pairs with successively smaller real part.


Let us stress the difference between these solutions and the Borel plane depicted in Fig.~\ref{fig:borelplots}. The Borel analysis reveals the exponential dependence i.e. the decay rate (for an exponential decay in $w$) and oscillation frequency (for oscillation in $w$). This analysis gives the subleading power law correction. For the transients, the exponential dependence is purely real, and one would be tempted to conclude that there is no oscillation. However, the imaginary parts of these solutions give rise to logarithmic oscillations as
\begin{equation}\label{eq:transientoscillation}
\Re(\sigma \tau^{\beta}) \propto \tau^{\Re(\beta)} \cos(\theta + \Im(\beta) \log(\tau)),
\end{equation}
for some phase $\theta$. We were unable to find other transients in the present setup and the fact that we nevertheless found an infinite set of modes, in principle capable to capture the whole information about the initial conditions, leads us to believe that there are not any. Let us repeat what is said in the caption of Fig.~\ref{fig:borelplots}. A generalization of the argument from Ref.~\cite{Heller:2016rtz} shows that the other exponents that can be read off from Fig.~\ref{fig:borelplots} are not physical modes.

Finally, we note that the argument presented above generalizes in a simple manner to the case of arbitrary~$\Delta$. In such a situation, the power law also gets contributions from $D(\tau,\tau_{0})$. To leading order in $w=\tau/\trel$, transient contributions to $\ed(\tau)$ behave as 
\begin{equation}\label{eq:deltatransient}
e^{-\frac{w}{1-\Delta/3}}w^{\beta +\frac{4\Delta}{45(1-\Delta/3)^2}},
\end{equation}
where $\beta$ satisfies $M(\beta(1-\Delta/3)-\Delta/3)=0$. This is the main result of this paper. It should be compared with what is found in holographic setups, where transients behave as in Eq.~\eqref{eq.QNMs}. There are three main differences. The first is the appearance of singularities in the Borel plane that do not represent transients on top of the hydro part. Second, the transients that do carry information are all stacked on top of each other in the Borel plane. This corresponds to identical exponential decay but with different power laws. Lastly, while in holography the transients generically oscillate in proper time, in this kinetic theory they do so in logarithmic time.

In Sec.~6 we corroborate these results with numerical solutions. 

\vspace{12 pt}

\noindent \emph{5. Resurgence and initial conditions for constant $\tau_{rel}$.--}
When $\trel$ is constant, Eq.~\eqref{eq.Baym} is linear. This is a great simplification, allowing us to investigate resurgent relations between the hydrodynamic and the non-hydrodynamic modes, as well as describe how to match initial data to amplitudes of transients. 

We start the resurgent analysis by calculating the coefficients in the power series $\ed_{\beta}$. It satisfies 
\begin{equation}
\ed_{\beta}(\tau) = \frac{\tau}{2\, \trel} \int_{0}^1 H\left(x\right) \ed_{\beta}(\tau x) \intd x.
\end{equation}
With a power series ansatz as in Eq.~\eqref{eq:transpower}, we can match powers and solve for the coefficients in the series. They satisfy the recursive equation
\begin{equation}
e_{\beta,k+1}=\frac{e_{\beta,k}}{M(\beta-k-1)}.
\end{equation}
An immediate question arises: What is the large order behaviour of $e_{\beta,k}$? Is it divergent and if so, will it tell us about additional transient modes? For large $k$,
\begin{equation}
\frac{e_{\beta,k+1}}{e_{\beta,k}} = -k + \left(\beta +\frac{4}{3}\right) + \frac{16}{45 k} +\dots.
\end{equation}
This can be turned into a differential equation and a solution of this equation is a function that at large $w$ behaves as
\begin{equation}
\label{eq.hydrofromtrans}
e^{w} w^{-\beta - 4/3 }\left(1 - \frac{16}{45 \, w} - \frac{424}{14175 \, w^2} \dots \right).
\end{equation}
To find the contribution to $\ed(\tau)$ we must into account $D(\tau,\tau_0)$ and $w^\beta$ in Eqs.~\eqref{eq:transientAnsatz} and \eqref{eq:transpower}. These cancel out the exponential and the $w^{-\beta}$ respectively, leaving us with a series whose leading power is $-4/3$. Given that in the current case of $\Delta = 0$, $w \sim \tau$, one immediately recognizes in it the famous Bjorken perfect fluid solution~\cite{Bjorken:1982qr}. By the use of Eq.~\eqref{eq.PLandPT} and \eqref{eq.defA}, one can calculate the corresponding series for $\pa$. This turns out to be 
\begin{equation}
\frac{8}{5 \,w}+\frac{88}{105\, w^2} + \dots
\end{equation}
Comparing with Eq.~\eqref{eq:Acoeffs}, and setting there $\Delta$ to 0, we see that this is in fact the hydrodynamic gradient expansion. Note that this argument holds for every value of allowed~$\beta$.

This is an explicit demonstration of resurgent properties of these solutions, see Refs.~\cite{Dunne,Aniceto:2018bis} for introductions to resurgence and Ref.~\cite{Aniceto:2015rua} for another example of resurgent phenomena in the context of integral equations. The gradient expansion can be reconstructed from the large order behavior of the transient, since in the constant relaxation case the only exponential contribution to ${\cal E}(\tau)$ with respect to each transient is the hydrodynamic series itself.

Now we describe how to map between initial conditions, described by $\ed_0$, and transient modes. This procedure only works for $\Delta=0$, i.e. when the problem is linear. Given a solution $\ed(\tau)$, one can trivially solve for $\ed_0(\tau)$ in Eq.~\eqref{eq.Baym}. Knowing the form of transients, one can calculate the corresponding $\ed_0$ to each transient. Thus, a decomposition of a solution $\ed$ into transients can be translated into a decomposition of $\ed_0$.

As a check of this, we have numerically calculated the $\ed_0$'s corresponding to the first two transients and compared these with the late time expansion of $\ed_0$ in Eq.~\eqref{eq:eps0series}. The characteristic features of this expansion, namely the leading power of $1/\tau$ and a vanishing quadratic term, can be verified for these solutions. 

\vspace{12 pt}

\noindent \emph{6. Comparison with numerical solutions.--}
In previous sections, we have calculated a family of transient modes, each exponentially decaying with the same rate but with different power laws. These powers were determined from a rather high-level argument and additional checks are required to be confident that they are physical modes. Indeed, as observed in Ref.~\cite{Heller:2016rtz}, the analytic structure of $H$ can give rise to unphysical modes. This section presents numerical evidence that they are physical. 

Our interest in looking at transients prompts the need for very precise numerics. We need a time interval long enough so that they are clearly separated from each other in magnitude. In holographic setups, the ratio of the magnitudes of the transients is exponentially large. In this case, there is only a power law suppression. Thus, this setup requires a longer interval of time compared to what a similar calculation in holography would need. Since the transients decay exponentially fast compared to the hydrodynamic contribution, this presents an obvious numerical challenge. Finite difference methods have an error that scales polynomially in the grid spacing which makes them unsuitable for studying exponentially small effects. More appropriate are spectral and pseudo-spectral methods which have an error that (for smooth functions) scales exponentially in the grid spacing. See, e.g., Refs.~\cite{Dutykh:2016spectral,boyd2001chebyshev} for introductions to these methods.

\begin{figure}[t]
	\centering
	\includegraphics[width=1\linewidth]{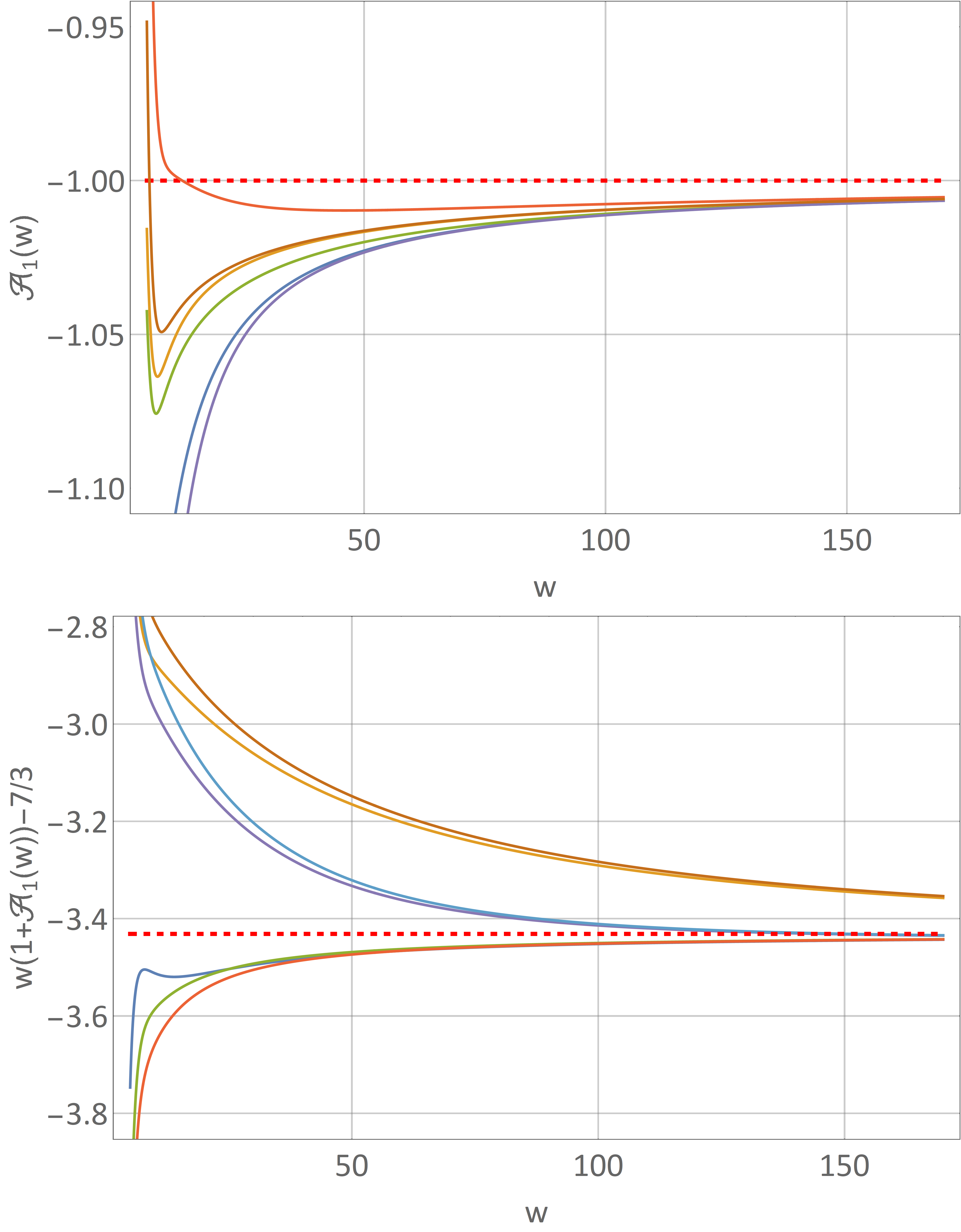}
	\caption{Shown here is $\pa_1(w)$ defined in Eq.~\eqref{eq:subtractions}. The plots provide overwhelming evidence that Eq~\eqref{eq:transient1} accurately describes the first transient mode. Note that \mbox{$\Delta = 0$.} (Top) All curves approach the exponential decay rate of the transient modes -1. (Bottom) All curves approach the power law decay rate of the first transient mode $\beta_1$.}
	\label{fig:transient1}
\end{figure}

Given an initial distribution function, the integral equation~\eqref{eq.Baym} can be solved by iteration. Choice of initial distribution function is made so that $\ed_0(\tau)$ can be calculated analytically. For $\Delta=0$, we calculated solutions on an interval from $w=5$ to $w=170$. This means we need an accuracy of at least $e^{-170} \approx 10^{-74}$. We achieved this by performing calculations in Mathematica with 1350 grid points and precision 900, iterating the equation until the maximal relative error between subsequent iterations was less than $10^{-150}$. For this process to converge, spectral filtering was used, see Ref.~\cite{boyd2001chebyshev}. For each initial condition we required several hours of computations on a powerful desktop computer. By a process of subtracting solutions of different initial conditions, we are able to study transients.

\begin{figure}[t]
	\centering
	\includegraphics[width=1\linewidth]{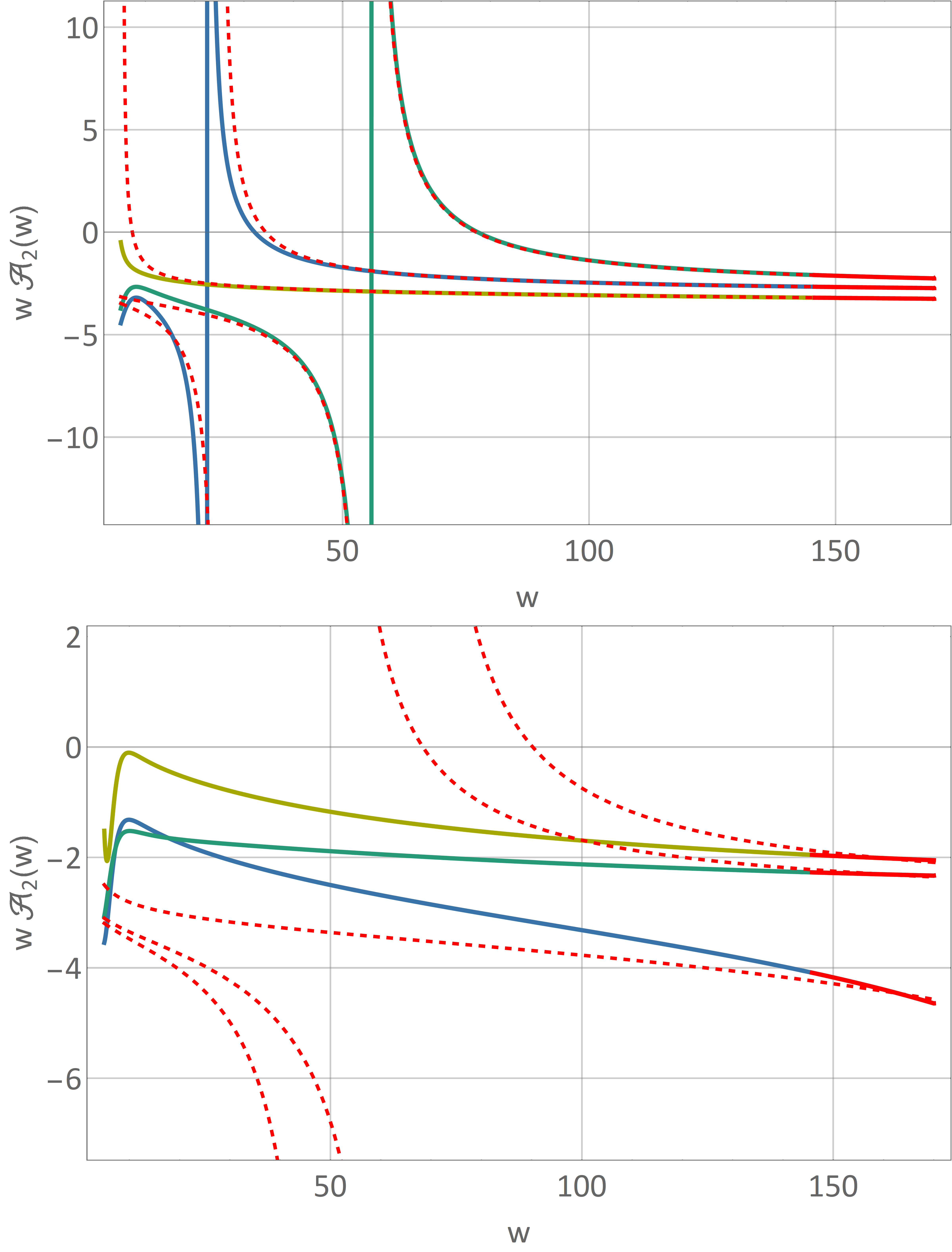}
    \caption{This figure compares numerical evaluated $\pa_2$, i.e, the second transient, with theoretical predictions for $\Delta=0$. Dashed red lines are of the form of Eq.~\eqref{eq:transientoscillation}, where $\theta$ is fitted using data in the continuous red colored region at late times. (Top) Eq~\eqref{eq:transientoscillation} describes the curves very well. Fitting also the value of $\beta_2$, it differs from the analytical value by less than $1\%$. The vertical segments represent singularity of the tangent function appearing in Eq.~\eqref{eq:betacurve}. (Bottom) Likely due to interference from subleading transients with large amplitudes, the fit does not work well.}
	\label{fig:beta2fits}
\end{figure}

Independent of initial conditions, $\pa$ behaves universally at late times, corresponding to the hydrodynamic gradient expansion, see Eq.~\eqref{eq:Acoeffs}. Subtracting two solutions will remove the universal behavior and leave only the transient behavior. Taking also a logarithmic derivative will remove the overall amplitude and we are left with a universal late time behavior corresponding to the transient mode. By taking into account the phase $\theta$ in Eq.~\eqref{eq:transientoscillation}, this subtraction can be repeated to get a sequence of functions whose behavior is universal at late times. Here, we consider the first two functions so obtained. These subtractions do not involve a phase, and so we define
\begin{align}\label{eq:subtractions}
\pa_{0} & = \pa\\
\pa_{1}  &= \frac{\intd}{\intd w}\log \left( \pa_{0} -\pa'_{0} \right) \\
\pa_{2}  &= \frac{\intd}{\intd w}\log \left( \pa_{1} -\pa'_{1} \right),
\end{align}
where the prime denotes solutions obtained using different initial conditions. $\pa_k$ will be related to the transient corresponding to $\beta_k$. Analytic calculation implies
\begin{equation}\label{eq:transient1}
\pa_1(w) = -1 + \frac{\beta_1+7/3}{w} + \dots,
\end{equation}
where the $-1$ comes from the exponential decay rate. As seen in Fig~\ref{fig:transient1}, both the decay rate and $\beta_1$ approach their predicted analytic values (red dashed line). 

The next transient is supposed to exhibit oscillations in logarithmic time. To leading order, $\pa_2$ satisfies
\begin{multline}\label{eq:betacurve}
w\pa_2(w) = \Re(\beta_2) - \beta_1 - 1 -\Im(\beta_2)  \tan \left(\theta +\Im(\beta_2) \log (w) \right) \\
\approx -3.0271  - 0.5614 \tan \left(\theta + 0.5614 \log (w) \right),
\end{multline}
where numerical values for $\beta_1$ and $\beta_2$ have been used. This has characteristic singularities that should have clear signals in the numerical solutions. However, corrections coming from subleading transients could spoil this if their amplitudes are large enough. Indeed, as Fig.~\ref{fig:beta2fits} shows,  some solutions are not well described by Eq.~\eqref{eq:betacurve} while others are so very well. With only one adjustable parameter and fitting only to a small interval at late times, one finds a remarkable agreement, see Fig.~\ref{fig:beta2fits}. This is striking confirmation of the multiplicity of cuts stacked on top of each other in the Borel plane and demonstrates the physicality of oscillations in logarithmic time. 

In addition, one can also fit $\beta_2$ to the data. The result matches the analytic value to better than $1\%$.

\vspace{12 pt}

\noindent \emph{7. Summary and outlook.--} In the present article we analyzed the non-hydrodynamic sector of kinetic theory in the relaxation time approximation. Using a three pronged approach involving asymptotic series, analytic solutions of an integral equation and high precision numerical solutions of initial value problem, we show how each of these methods allow us to probe this sector. The relaxation time was taken to exhibit general power law dependence on the effective temperature, see Eq.~\eqref{eq.taurel}. Such a theory was regarded here as a toy model of weakly-coupled gauge theory dynamics. Moreover, we focused on expanding plasma systems undergoing rapid longitudinal expansion, similarly to ultrarelativistic heavy ion collisions. We simplified our treatment by further assuming boost-invariance along the expansion axis and no transversal dynamics. Our chief motivation, inspired by similar analyses in holography, was to understand what imprint weakly-coupled transient effects will have on the energy-momentum tensor of expanding plasma. The fact that the Boltzmann equation requires for solving the initial value problem specifying a function not only of spacetime coordinates but also of momenta indicated that there should be infinitely many transient effects carrying information about a given initial condition to late times. This intuition turned out to be correct and we discovered that in the expanding plasma system in RTA kinetic theory there are infinitely many exponentially suppressed contributions to the energy-momentum tensor, decaying on a timescale $\tau_\text{decay}=(1-\Delta/3)\trel$, see Sec.~4 and Eq.~\eqref{eq:deltatransient}. What differentiates these transients from each other is the subleading behavior. We show that it consists of different power law decay and oscillations in logarithmic time, see Eqs.~\eqref{eq:transientoscillation} and~\eqref{eq:deltatransient}, as well as Fig.~\ref{fig:betas}. We corroborate both findings with the analysis of large orders of the hydrodynamic gradient expansion, see Sec.~3 and Fig.~\ref{fig:borelplots}, and explicit solution of the initial value problem, see Sec.~6 and Figs.~\ref{fig:transient1} and \ref{fig:beta2fits}, noting very good agreement. The latter was achieved by a very accurate way of implementing the initial value problem given by the pseudospectral methods and use of the iterative scheme from Ref.~\cite{Florkowski:2013lya}. 

Furthermore, similarly to the studies reported in Ref.~\cite{Heller:2016rtz}, we see singularities of the Borel transform of the hydrodynamic gradient expansion that do not correspond to modes of the expanding plasma, see Fig.~\ref{fig:borelplots}. For $\Delta>3/2$, these would represent the dominant contribution to transient behavior in the initial value problem, something which we do not see. What is also interesting is that for $\Delta > 2$, these singularities become the dominant effects controlling the divergence of the hydrodynamic gradient expansion, as opposed to the least damped transients in all the other known setups dealing with hydrodynamics, see, e.g., Ref~\cite{Florkowski:2017olj} for a review. However, there are intriguing similarities with the so-called ghost-instantons explored in a quantum mechanical setting in Ref.~\cite{Basar:2013eka}. 

We note that for constant relaxation time ($\Delta = 0$) the integral equation for the energy density becomes linear, see Eq.~\eqref{eq.Baym}. Here we find beautiful resurgent relations in which large orders of the hydrodynamic gradient expansion carry information about the transient modes and the large order gradient expansion accompanying each transient mode is controlled by the hydrodynamic series, see Eq.~\eqref{eq.hydrofromtrans}. As a result, the trans-series ansatz in this case consists only of two types of contributions: the hydrodynamic series and a sum over transient modes without any further nonlinear effects. For $\Delta \neq 0$, we expect nonlinear effects. The question of what happens when $\Delta>3$ or at the point of breakdown, $\Delta=3$, is left open.

Our work raises several interesting questions. Perhaps the most important one is what kind of transients in expanding plasma systems (or other setups undergoing macroscopic motion) exists for other collisional kernels and are their decay rates comparable~/~the~same? This is of relevance in the search for transient effects in heavy-ion collision or cold atoms experiments, see, e.g., Refs.~\cite{Liu:2015nwa,Brewer:2015ipa}.

Another interesting question is if it is possible to derive the properties of the transients directly from singularities of the retarded two-point function of the energy-momentum tensor studied in Ref.~\cite{Romatschke:2015gic}. The reason why we expect such a link to exist is, first, that similar analysis works out in holography, see Ref.~\cite{Janik:2006gp}, and, second, that the properties of transients are related with the properties of the function $H(q)$ given by Eq.~\eqref{eq.defH} and the latter is related to properties of equilibrium, i.e. the equilibrium distribution function given by Eq.~\eqref{eq.feq}. Furthermore, the Green's function analysis in Ref.~\cite{Romatschke:2015gic} reveals branch cut singularities with the imaginary part of branch points, responsible for dissipation, being given by the inverse of the relaxation time and this is precisely what we observed here. Such a method of translating from the singularities of the energy-momentum tensor Green's functions to expanding plasma systems may shed light on how transient manifest themselves both for kinetic theories with more complicated collisional kernel, see recent Ref.~\cite{Kurkela:2017xis}, and for more general flows.

On the latter front, it would be very interesting to generalize the present analysis to other flows, starting from the most symmetric ones such as cosmological expansion addressed in Ref.~\cite{Bazow:2015dha,Buchel:2016cbj} or (perturbations of) the so-called Gubser flow~\cite{Gubser:2010ze} studied in the RTA kinetic theory in Refs.~\cite{Denicol:2014xca,Denicol:2014tha}. An interesting aspect for such a comparison is the question what happens when the relaxation time from the present setup scales with the effective temperature faster than $\frac{1}{T^{3}}$. For such relaxation times, we do not expect local equilibrium in the energy-momentum tensor at asymptotically late times and similar phenomenon was indeed seen in Refs.~\cite{Denicol:2014xca,Denicol:2014tha}. It would be, therefore, interesting to understand if in such cases hydrodynamics becomes a good description of the boost-invariant plasma for a window of intermediate times and how the system exits the hydrodynamic regime.

\vspace{10 pt}
\begin{acknowledgments}

We are grateful to our collaborators on~\cite{Heller:2016rtz}, A.~Kurkela and M.~Spalinski, and friends and colleagues, in particular I.~Aniceto, G.~Basar, J.~Casalderrey-Solana, G.~Dunne, W.~Florkowski, U.~Heinz, R.~Janik, P.~Kovtun, H.~Marrochio, M.~Martines, J. Noronha, P.~Romatschke, M.~Strickland, L.~Yaffe, and P.~Witaszczyk and for useful discussions, correspondence and comments on the draft. Gravity, Quantum Fields and Information group at Albert Einstein Institute that we are part of is generously supported by the Alexander von Humboldt Foundation and the Federal Ministry for Education and Research through the Sofja Kovalevskaja Award. V.S. also acknowledges partial support from the National Science
Centre through grant 2015/19/B/ST2/02824.

\end{acknowledgments}

\vspace{12 pt}

\appendix

\bibliography{hydromodes}{}

\begin{thebibliography}{64}%
\makeatletter
\providecommand \@ifxundefined [1]{%
 \@ifx{#1\undefined}
}%
\providecommand \@ifnum [1]{%
 \ifnum #1\expandafter \@firstoftwo
 \else \expandafter \@secondoftwo
 \fi
}%
\providecommand \@ifx [1]{%
 \ifx #1\expandafter \@firstoftwo
 \else \expandafter \@secondoftwo
 \fi
}%
\providecommand \natexlab [1]{#1}%
\providecommand \enquote  [1]{``#1''}%
\providecommand \bibnamefont  [1]{#1}%
\providecommand \bibfnamefont [1]{#1}%
\providecommand \citenamefont [1]{#1}%
\providecommand \href@noop [0]{\@secondoftwo}%
\providecommand \href [0]{\begingroup \@sanitize@url \@href}%
\providecommand \@href[1]{\@@startlink{#1}\@@href}%
\providecommand \@@href[1]{\endgroup#1\@@endlink}%
\providecommand \@sanitize@url [0]{\catcode `\\12\catcode `\$12\catcode
  `\&12\catcode `\#12\catcode `\^12\catcode `\_12\catcode `\%12\relax}%
\providecommand \@@startlink[1]{}%
\providecommand \@@endlink[0]{}%
\providecommand \url  [0]{\begingroup\@sanitize@url \@url }%
\providecommand \@url [1]{\endgroup\@href {#1}{\urlprefix }}%
\providecommand \urlprefix  [0]{URL }%
\providecommand \Eprint [0]{\href }%
\providecommand \doibase [0]{http://dx.doi.org/}%
\providecommand \selectlanguage [0]{\@gobble}%
\providecommand \bibinfo  [0]{\@secondoftwo}%
\providecommand \bibfield  [0]{\@secondoftwo}%
\providecommand \translation [1]{[#1]}%
\providecommand \BibitemOpen [0]{}%
\providecommand \bibitemStop [0]{}%
\providecommand \bibitemNoStop [0]{.\EOS\space}%
\providecommand \EOS [0]{\spacefactor3000\relax}%
\providecommand \BibitemShut  [1]{\csname bibitem#1\endcsname}%
\let\auto@bib@innerbib\@empty
\bibitem [{\citenamefont {Heinz}(2005)}]{Heinz:2004pj}%
  \BibitemOpen
  \bibfield  {author} {\bibinfo {author} {\bibfnamefont {U.~W.}\ \bibnamefont
  {Heinz}},\ }\href {\doibase 10.1063/1.1843595} {\bibfield  {journal}
  {\bibinfo  {journal} {AIP Conf.Proc.}\ }\textbf {\bibinfo {volume} {739}},\
  \bibinfo {pages} {163} (\bibinfo {year} {2005})},\ \Eprint
  {http://arxiv.org/abs/nucl-th/0407067} {arXiv:nucl-th/0407067 [nucl-th]}
  \BibitemShut {NoStop}%
\bibitem [{\citenamefont {Casalderrey-Solana}\ \emph
  {et~al.}(2011)\citenamefont {Casalderrey-Solana}, \citenamefont {Liu},
  \citenamefont {Mateos}, \citenamefont {Rajagopal},\ and\ \citenamefont
  {Wiedemann}}]{CasalderreySolana:2011us}%
  \BibitemOpen
  \bibfield  {author} {\bibinfo {author} {\bibfnamefont {J.}~\bibnamefont
  {Casalderrey-Solana}}, \bibinfo {author} {\bibfnamefont {H.}~\bibnamefont
  {Liu}}, \bibinfo {author} {\bibfnamefont {D.}~\bibnamefont {Mateos}},
  \bibinfo {author} {\bibfnamefont {K.}~\bibnamefont {Rajagopal}}, \ and\
  \bibinfo {author} {\bibfnamefont {U.~A.}\ \bibnamefont {Wiedemann}},\ }\href
  {\doibase 10.1017/CBO9781139136747} {\  (\bibinfo {year} {2011}),\
  10.1017/CBO9781139136747},\ \Eprint {http://arxiv.org/abs/1101.0618}
  {arXiv:1101.0618 [hep-th]} \BibitemShut {NoStop}%
\bibitem [{\citenamefont {Busza}\ \emph {et~al.}(2018)\citenamefont {Busza},
  \citenamefont {Rajagopal},\ and\ \citenamefont {van~der
  Schee}}]{Busza:2018rrf}%
  \BibitemOpen
  \bibfield  {author} {\bibinfo {author} {\bibfnamefont {W.}~\bibnamefont
  {Busza}}, \bibinfo {author} {\bibfnamefont {K.}~\bibnamefont {Rajagopal}}, \
  and\ \bibinfo {author} {\bibfnamefont {W.}~\bibnamefont {van~der Schee}},\
  }\href@noop {} {\  (\bibinfo {year} {2018})},\ \Eprint
  {http://arxiv.org/abs/1802.04801} {arXiv:1802.04801 [hep-ph]} \BibitemShut
  {NoStop}%
\bibitem [{\citenamefont {Maldacena}(1998)}]{Maldacena:1997re}%
  \BibitemOpen
  \bibfield  {author} {\bibinfo {author} {\bibfnamefont {J.~M.}\ \bibnamefont
  {Maldacena}},\ }\href@noop {} {\bibfield  {journal} {\bibinfo  {journal}
  {Adv.Theor.Math.Phys.}\ }\textbf {\bibinfo {volume} {2}},\ \bibinfo {pages}
  {231} (\bibinfo {year} {1998})},\ \Eprint
  {http://arxiv.org/abs/hep-th/9711200} {arXiv:hep-th/9711200 [hep-th]}
  \BibitemShut {NoStop}%
\bibitem [{\citenamefont {Witten}(1998)}]{Witten:1998qj}%
  \BibitemOpen
  \bibfield  {author} {\bibinfo {author} {\bibfnamefont {E.}~\bibnamefont
  {Witten}},\ }\href@noop {} {\bibfield  {journal} {\bibinfo  {journal}
  {Adv.Theor.Math.Phys.}\ }\textbf {\bibinfo {volume} {2}},\ \bibinfo {pages}
  {253} (\bibinfo {year} {1998})},\ \Eprint
  {http://arxiv.org/abs/hep-th/9802150} {arXiv:hep-th/9802150 [hep-th]}
  \BibitemShut {NoStop}%
\bibitem [{\citenamefont {Gubser}\ \emph {et~al.}(1998)\citenamefont {Gubser},
  \citenamefont {Klebanov},\ and\ \citenamefont {Polyakov}}]{Gubser:1998bc}%
  \BibitemOpen
  \bibfield  {author} {\bibinfo {author} {\bibfnamefont {S.}~\bibnamefont
  {Gubser}}, \bibinfo {author} {\bibfnamefont {I.~R.}\ \bibnamefont
  {Klebanov}}, \ and\ \bibinfo {author} {\bibfnamefont {A.~M.}\ \bibnamefont
  {Polyakov}},\ }\href {\doibase 10.1016/S0370-2693(98)00377-3} {\bibfield
  {journal} {\bibinfo  {journal} {Phys.Lett.}\ }\textbf {\bibinfo {volume}
  {B428}},\ \bibinfo {pages} {105} (\bibinfo {year} {1998})},\ \Eprint
  {http://arxiv.org/abs/hep-th/9802109} {arXiv:hep-th/9802109 [hep-th]}
  \BibitemShut {NoStop}%
\bibitem [{\citenamefont {Florkowski}\ \emph {et~al.}(2018)\citenamefont
  {Florkowski}, \citenamefont {Heller},\ and\ \citenamefont
  {Spalinski}}]{Florkowski:2017olj}%
  \BibitemOpen
  \bibfield  {author} {\bibinfo {author} {\bibfnamefont {W.}~\bibnamefont
  {Florkowski}}, \bibinfo {author} {\bibfnamefont {M.~P.}\ \bibnamefont
  {Heller}}, \ and\ \bibinfo {author} {\bibfnamefont {M.}~\bibnamefont
  {Spalinski}},\ }\href {\doibase 10.1088/1361-6633/aaa091} {\bibfield
  {journal} {\bibinfo  {journal} {Rept. Prog. Phys.}\ }\textbf {\bibinfo
  {volume} {81}},\ \bibinfo {pages} {046001} (\bibinfo {year} {2018})},\
  \Eprint {http://arxiv.org/abs/1707.02282} {arXiv:1707.02282 [hep-ph]}
  \BibitemShut {NoStop}%
\bibitem [{\citenamefont
  {Romatschke}(2017{\natexlab{a}})}]{Romatschke:2017vte}%
  \BibitemOpen
  \bibfield  {author} {\bibinfo {author} {\bibfnamefont {P.}~\bibnamefont
  {Romatschke}},\ }\href@noop {} {\  (\bibinfo {year} {2017}{\natexlab{a}})},\
  \Eprint {http://arxiv.org/abs/1704.08699} {arXiv:1704.08699 [hep-th]}
  \BibitemShut {NoStop}%
\bibitem [{\citenamefont {Bjorken}(1983)}]{Bjorken:1982qr}%
  \BibitemOpen
  \bibfield  {author} {\bibinfo {author} {\bibfnamefont {J.}~\bibnamefont
  {Bjorken}},\ }\href {\doibase 10.1103/PhysRevD.27.140} {\bibfield  {journal}
  {\bibinfo  {journal} {Phys.Rev.}\ }\textbf {\bibinfo {volume} {D27}},\
  \bibinfo {pages} {140} (\bibinfo {year} {1983})}\BibitemShut {NoStop}%
\bibitem [{\citenamefont {Chesler}\ and\ \citenamefont
  {Yaffe}(2010)}]{Chesler:2009cy}%
  \BibitemOpen
  \bibfield  {author} {\bibinfo {author} {\bibfnamefont {P.~M.}\ \bibnamefont
  {Chesler}}\ and\ \bibinfo {author} {\bibfnamefont {L.~G.}\ \bibnamefont
  {Yaffe}},\ }\href {\doibase 10.1103/PhysRevD.82.026006} {\bibfield  {journal}
  {\bibinfo  {journal} {Phys.Rev.}\ }\textbf {\bibinfo {volume} {D82}},\
  \bibinfo {pages} {026006} (\bibinfo {year} {2010})},\ \Eprint
  {http://arxiv.org/abs/0906.4426} {arXiv:0906.4426 [hep-th]} \BibitemShut
  {NoStop}%
\bibitem [{\citenamefont {Janik}\ and\ \citenamefont
  {Peschanski}(2006)}]{Janik:2006gp}%
  \BibitemOpen
  \bibfield  {author} {\bibinfo {author} {\bibfnamefont {R.~A.}\ \bibnamefont
  {Janik}}\ and\ \bibinfo {author} {\bibfnamefont {R.~B.}\ \bibnamefont
  {Peschanski}},\ }\href {\doibase 10.1103/PhysRevD.74.046007} {\bibfield
  {journal} {\bibinfo  {journal} {Phys. Rev.}\ }\textbf {\bibinfo {volume}
  {D74}},\ \bibinfo {pages} {046007} (\bibinfo {year} {2006})},\ \Eprint
  {http://arxiv.org/abs/hep-th/0606149} {arXiv:hep-th/0606149 [hep-th]}
  \BibitemShut {NoStop}%
\bibitem [{\citenamefont {Heller}\ \emph {et~al.}(2014)\citenamefont {Heller},
  \citenamefont {Janik}, \citenamefont {Spaliński},\ and\ \citenamefont
  {Witaszczyk}}]{Heller:2014wfa}%
  \BibitemOpen
  \bibfield  {author} {\bibinfo {author} {\bibfnamefont {M.~P.}\ \bibnamefont
  {Heller}}, \bibinfo {author} {\bibfnamefont {R.~A.}\ \bibnamefont {Janik}},
  \bibinfo {author} {\bibfnamefont {M.}~\bibnamefont {Spaliński}}, \ and\
  \bibinfo {author} {\bibfnamefont {P.}~\bibnamefont {Witaszczyk}},\ }\href
  {\doibase 10.1103/PhysRevLett.113.261601} {\bibfield  {journal} {\bibinfo
  {journal} {Phys.Rev.Lett.}\ }\textbf {\bibinfo {volume} {113}},\ \bibinfo
  {pages} {261601} (\bibinfo {year} {2014})},\ \Eprint
  {http://arxiv.org/abs/1409.5087} {arXiv:1409.5087 [hep-th]} \BibitemShut
  {NoStop}%
\bibitem [{\citenamefont {Hartnoll}\ \emph {et~al.}(2016)\citenamefont
  {Hartnoll}, \citenamefont {Lucas},\ and\ \citenamefont
  {Sachdev}}]{Hartnoll:2016apf}%
  \BibitemOpen
  \bibfield  {author} {\bibinfo {author} {\bibfnamefont {S.~A.}\ \bibnamefont
  {Hartnoll}}, \bibinfo {author} {\bibfnamefont {A.}~\bibnamefont {Lucas}}, \
  and\ \bibinfo {author} {\bibfnamefont {S.}~\bibnamefont {Sachdev}},\
  }\href@noop {} {\  (\bibinfo {year} {2016})},\ \Eprint
  {http://arxiv.org/abs/1612.07324} {arXiv:1612.07324 [hep-th]} \BibitemShut
  {NoStop}%
\bibitem [{\citenamefont {Kovtun}\ and\ \citenamefont
  {Starinets}(2005)}]{Kovtun:2005ev}%
  \BibitemOpen
  \bibfield  {author} {\bibinfo {author} {\bibfnamefont {P.~K.}\ \bibnamefont
  {Kovtun}}\ and\ \bibinfo {author} {\bibfnamefont {A.~O.}\ \bibnamefont
  {Starinets}},\ }\href {\doibase 10.1103/PhysRevD.72.086009} {\bibfield
  {journal} {\bibinfo  {journal} {Phys.Rev.}\ }\textbf {\bibinfo {volume}
  {D72}},\ \bibinfo {pages} {086009} (\bibinfo {year} {2005})},\ \Eprint
  {http://arxiv.org/abs/hep-th/0506184} {arXiv:hep-th/0506184 [hep-th]}
  \BibitemShut {NoStop}%
\bibitem [{\citenamefont {Romatschke}(2016)}]{Romatschke:2015gic}%
  \BibitemOpen
  \bibfield  {author} {\bibinfo {author} {\bibfnamefont {P.}~\bibnamefont
  {Romatschke}},\ }\href {\doibase 10.1140/epjc/s10052-016-4169-7} {\bibfield
  {journal} {\bibinfo  {journal} {Eur. Phys. J.}\ }\textbf {\bibinfo {volume}
  {C76}},\ \bibinfo {pages} {352} (\bibinfo {year} {2016})},\ \Eprint
  {http://arxiv.org/abs/1512.02641} {arXiv:1512.02641 [hep-th]} \BibitemShut
  {NoStop}%
\bibitem [{\citenamefont {Kurkela}\ and\ \citenamefont
  {Wiedemann}(2017)}]{Kurkela:2017xis}%
  \BibitemOpen
  \bibfield  {author} {\bibinfo {author} {\bibfnamefont {A.}~\bibnamefont
  {Kurkela}}\ and\ \bibinfo {author} {\bibfnamefont {U.~A.}\ \bibnamefont
  {Wiedemann}},\ }\href@noop {} {\  (\bibinfo {year} {2017})},\ \Eprint
  {http://arxiv.org/abs/1712.04376} {arXiv:1712.04376 [hep-ph]} \BibitemShut
  {NoStop}%
\bibitem [{\citenamefont {Denicol}\ and\ \citenamefont
  {Noronha}(2016)}]{Denicol:2016bjh}%
  \BibitemOpen
  \bibfield  {author} {\bibinfo {author} {\bibfnamefont {G.~S.}\ \bibnamefont
  {Denicol}}\ and\ \bibinfo {author} {\bibfnamefont {J.}~\bibnamefont
  {Noronha}},\ }\href@noop {} {\  (\bibinfo {year} {2016})},\ \Eprint
  {http://arxiv.org/abs/1608.07869} {arXiv:1608.07869 [nucl-th]} \BibitemShut
  {NoStop}%
\bibitem [{\citenamefont {Heller}\ \emph {et~al.}(2018)\citenamefont {Heller},
  \citenamefont {Kurkela}, \citenamefont {Spalinski},\ and\ \citenamefont
  {Svensson}}]{Heller:2016rtz}%
  \BibitemOpen
  \bibfield  {author} {\bibinfo {author} {\bibfnamefont {M.~P.}\ \bibnamefont
  {Heller}}, \bibinfo {author} {\bibfnamefont {A.}~\bibnamefont {Kurkela}},
  \bibinfo {author} {\bibfnamefont {M.}~\bibnamefont {Spalinski}}, \ and\
  \bibinfo {author} {\bibfnamefont {V.}~\bibnamefont {Svensson}},\ }\href
  {\doibase 10.1103/PhysRevD.97.091503} {\bibfield  {journal} {\bibinfo
  {journal} {Phys. Rev.}\ }\textbf {\bibinfo {volume} {D97}},\ \bibinfo {pages}
  {091503} (\bibinfo {year} {2018})},\ \Eprint
  {http://arxiv.org/abs/1609.04803} {arXiv:1609.04803 [nucl-th]} \BibitemShut
  {NoStop}%
\bibitem [{\citenamefont {Blaizot}\ and\ \citenamefont
  {Yan}(2017{\natexlab{a}})}]{Blaizot:2017lht}%
  \BibitemOpen
  \bibfield  {author} {\bibinfo {author} {\bibfnamefont {J.-P.}\ \bibnamefont
  {Blaizot}}\ and\ \bibinfo {author} {\bibfnamefont {L.}~\bibnamefont {Yan}},\
  }\href {\doibase 10.1007/JHEP11(2017)161} {\bibfield  {journal} {\bibinfo
  {journal} {JHEP}\ }\textbf {\bibinfo {volume} {11}},\ \bibinfo {pages} {161}
  (\bibinfo {year} {2017}{\natexlab{a}})},\ \Eprint
  {http://arxiv.org/abs/1703.10694} {arXiv:1703.10694 [nucl-th]} \BibitemShut
  {NoStop}%
\bibitem [{\citenamefont {Blaizot}\ and\ \citenamefont
  {Yan}(2017{\natexlab{b}})}]{Blaizot:2017ucy}%
  \BibitemOpen
  \bibfield  {author} {\bibinfo {author} {\bibfnamefont {J.-P.}\ \bibnamefont
  {Blaizot}}\ and\ \bibinfo {author} {\bibfnamefont {L.}~\bibnamefont {Yan}},\
  }\href@noop {} {\  (\bibinfo {year} {2017}{\natexlab{b}})},\ \Eprint
  {http://arxiv.org/abs/1712.03856} {arXiv:1712.03856 [nucl-th]} \BibitemShut
  {NoStop}%
\bibitem [{\citenamefont {Baier}\ \emph {et~al.}(2008)\citenamefont {Baier},
  \citenamefont {Romatschke}, \citenamefont {Son}, \citenamefont {Starinets},\
  and\ \citenamefont {Stephanov}}]{Baier:2007ix}%
  \BibitemOpen
  \bibfield  {author} {\bibinfo {author} {\bibfnamefont {R.}~\bibnamefont
  {Baier}}, \bibinfo {author} {\bibfnamefont {P.}~\bibnamefont {Romatschke}},
  \bibinfo {author} {\bibfnamefont {D.~T.}\ \bibnamefont {Son}}, \bibinfo
  {author} {\bibfnamefont {A.~O.}\ \bibnamefont {Starinets}}, \ and\ \bibinfo
  {author} {\bibfnamefont {M.~A.}\ \bibnamefont {Stephanov}},\ }\href {\doibase
  10.1088/1126-6708/2008/04/100} {\bibfield  {journal} {\bibinfo  {journal}
  {JHEP}\ }\textbf {\bibinfo {volume} {04}},\ \bibinfo {pages} {100} (\bibinfo
  {year} {2008})},\ \Eprint {http://arxiv.org/abs/0712.2451} {arXiv:0712.2451
  [hep-th]} \BibitemShut {NoStop}%
\bibitem [{\citenamefont {Florkowski}\ and\ \citenamefont
  {Ryblewski}(2011)}]{Florkowski:2010cf}%
  \BibitemOpen
  \bibfield  {author} {\bibinfo {author} {\bibfnamefont {W.}~\bibnamefont
  {Florkowski}}\ and\ \bibinfo {author} {\bibfnamefont {R.}~\bibnamefont
  {Ryblewski}},\ }\href {\doibase 10.1103/PhysRevC.83.034907} {\bibfield
  {journal} {\bibinfo  {journal} {Phys. Rev.}\ }\textbf {\bibinfo {volume}
  {C83}},\ \bibinfo {pages} {034907} (\bibinfo {year} {2011})},\ \Eprint
  {http://arxiv.org/abs/1007.0130} {arXiv:1007.0130 [nucl-th]} \BibitemShut
  {NoStop}%
\bibitem [{\citenamefont {Martinez}\ and\ \citenamefont
  {Strickland}(2010)}]{Martinez:2010sc}%
  \BibitemOpen
  \bibfield  {author} {\bibinfo {author} {\bibfnamefont {M.}~\bibnamefont
  {Martinez}}\ and\ \bibinfo {author} {\bibfnamefont {M.}~\bibnamefont
  {Strickland}},\ }\href {\doibase 10.1016/j.nuclphysa.2010.08.011} {\bibfield
  {journal} {\bibinfo  {journal} {Nucl. Phys.}\ }\textbf {\bibinfo {volume}
  {A848}},\ \bibinfo {pages} {183} (\bibinfo {year} {2010})},\ \Eprint
  {http://arxiv.org/abs/1007.0889} {arXiv:1007.0889 [nucl-th]} \BibitemShut
  {NoStop}%
\bibitem [{\citenamefont {Denicol}\ \emph
  {et~al.}(2012{\natexlab{a}})\citenamefont {Denicol}, \citenamefont {Niemi},
  \citenamefont {Molnar},\ and\ \citenamefont {Rischke}}]{Denicol:2012cn}%
  \BibitemOpen
  \bibfield  {author} {\bibinfo {author} {\bibfnamefont {G.~S.}\ \bibnamefont
  {Denicol}}, \bibinfo {author} {\bibfnamefont {H.}~\bibnamefont {Niemi}},
  \bibinfo {author} {\bibfnamefont {E.}~\bibnamefont {Molnar}}, \ and\ \bibinfo
  {author} {\bibfnamefont {D.~H.}\ \bibnamefont {Rischke}},\ }\href {\doibase
  10.1103/PhysRevD.85.114047, 10.1103/PhysRevD.91.039902} {\bibfield  {journal}
  {\bibinfo  {journal} {Phys. Rev.}\ }\textbf {\bibinfo {volume} {D85}},\
  \bibinfo {pages} {114047} (\bibinfo {year} {2012}{\natexlab{a}})},\ \bibinfo
  {note} {[Erratum: Phys. Rev.D91,no.3,039902(2015)]},\ \Eprint
  {http://arxiv.org/abs/1202.4551} {arXiv:1202.4551 [nucl-th]} \BibitemShut
  {NoStop}%
\bibitem [{\citenamefont {Denicol}\ \emph
  {et~al.}(2012{\natexlab{b}})\citenamefont {Denicol}, \citenamefont {Molnar},
  \citenamefont {Niemi},\ and\ \citenamefont {Rischke}}]{Denicol:2012es}%
  \BibitemOpen
  \bibfield  {author} {\bibinfo {author} {\bibfnamefont {G.}~\bibnamefont
  {Denicol}}, \bibinfo {author} {\bibfnamefont {E.}~\bibnamefont {Molnar}},
  \bibinfo {author} {\bibfnamefont {H.}~\bibnamefont {Niemi}}, \ and\ \bibinfo
  {author} {\bibfnamefont {D.}~\bibnamefont {Rischke}},\ }\href {\doibase
  10.1140/epja/i2012-12170-x} {\bibfield  {journal} {\bibinfo  {journal} {Eur.
  Phys. J. A}\ }\textbf {\bibinfo {volume} {48}},\ \bibinfo {pages} {170}
  (\bibinfo {year} {2012}{\natexlab{b}})},\ \Eprint
  {http://arxiv.org/abs/1206.1554} {arXiv:1206.1554 [nucl-th]} \BibitemShut
  {NoStop}%
\bibitem [{\citenamefont {Stephanov}\ and\ \citenamefont
  {Yin}(2017)}]{Stephanov:2017ghc}%
  \BibitemOpen
  \bibfield  {author} {\bibinfo {author} {\bibfnamefont {M.}~\bibnamefont
  {Stephanov}}\ and\ \bibinfo {author} {\bibfnamefont {Y.}~\bibnamefont
  {Yin}},\ }\href@noop {} {\  (\bibinfo {year} {2017})},\ \Eprint
  {http://arxiv.org/abs/1712.10305} {arXiv:1712.10305 [nucl-th]} \BibitemShut
  {NoStop}%
\bibitem [{\citenamefont {Heller}\ \emph {et~al.}(2013)\citenamefont {Heller},
  \citenamefont {Janik},\ and\ \citenamefont {Witaszczyk}}]{Heller:2013fn}%
  \BibitemOpen
  \bibfield  {author} {\bibinfo {author} {\bibfnamefont {M.~P.}\ \bibnamefont
  {Heller}}, \bibinfo {author} {\bibfnamefont {R.~A.}\ \bibnamefont {Janik}}, \
  and\ \bibinfo {author} {\bibfnamefont {P.}~\bibnamefont {Witaszczyk}},\
  }\href {\doibase 10.1103/PhysRevLett.110.211602} {\bibfield  {journal}
  {\bibinfo  {journal} {Phys.Rev.Lett.}\ }\textbf {\bibinfo {volume} {110}},\
  \bibinfo {pages} {211602} (\bibinfo {year} {2013})},\ \Eprint
  {http://arxiv.org/abs/1302.0697} {arXiv:1302.0697 [hep-th]} \BibitemShut
  {NoStop}%
\bibitem [{\citenamefont {Heller}\ and\ \citenamefont
  {Spalinski}(2015)}]{Heller:2015dha}%
  \BibitemOpen
  \bibfield  {author} {\bibinfo {author} {\bibfnamefont {M.~P.}\ \bibnamefont
  {Heller}}\ and\ \bibinfo {author} {\bibfnamefont {M.}~\bibnamefont
  {Spalinski}},\ }\href {\doibase 10.1103/PhysRevLett.115.072501} {\bibfield
  {journal} {\bibinfo  {journal} {Phys. Rev. Lett.}\ }\textbf {\bibinfo
  {volume} {115}},\ \bibinfo {pages} {072501} (\bibinfo {year} {2015})},\
  \Eprint {http://arxiv.org/abs/1503.07514} {arXiv:1503.07514 [hep-th]}
  \BibitemShut {NoStop}%
\bibitem [{\citenamefont {Basar}\ and\ \citenamefont
  {Dunne}(2015)}]{Basar:2015ava}%
  \BibitemOpen
  \bibfield  {author} {\bibinfo {author} {\bibfnamefont {G.}~\bibnamefont
  {Basar}}\ and\ \bibinfo {author} {\bibfnamefont {G.~V.}\ \bibnamefont
  {Dunne}},\ }\href {\doibase 10.1103/PhysRevD.92.125011} {\bibfield  {journal}
  {\bibinfo  {journal} {Phys. Rev.}\ }\textbf {\bibinfo {volume} {D92}},\
  \bibinfo {pages} {125011} (\bibinfo {year} {2015})},\ \Eprint
  {http://arxiv.org/abs/1509.05046} {arXiv:1509.05046 [hep-th]} \BibitemShut
  {NoStop}%
\bibitem [{\citenamefont {Aniceto}\ and\ \citenamefont
  {Spalinski}(2016)}]{Aniceto:2015mto}%
  \BibitemOpen
  \bibfield  {author} {\bibinfo {author} {\bibfnamefont {I.}~\bibnamefont
  {Aniceto}}\ and\ \bibinfo {author} {\bibfnamefont {M.}~\bibnamefont
  {Spalinski}},\ }\href {\doibase 10.1103/PhysRevD.93.085008} {\bibfield
  {journal} {\bibinfo  {journal} {Phys. Rev.}\ }\textbf {\bibinfo {volume}
  {D93}},\ \bibinfo {pages} {085008} (\bibinfo {year} {2016})},\ \Eprint
  {http://arxiv.org/abs/1511.06358} {arXiv:1511.06358 [hep-th]} \BibitemShut
  {NoStop}%
\bibitem [{\citenamefont {Spalinski}(2018)}]{Spalinski:2017mel}%
  \BibitemOpen
  \bibfield  {author} {\bibinfo {author} {\bibfnamefont {M.}~\bibnamefont
  {Spalinski}},\ }\href {\doibase 10.1016/j.physletb.2017.11.059} {\bibfield
  {journal} {\bibinfo  {journal} {Phys. Lett.}\ }\textbf {\bibinfo {volume}
  {B776}},\ \bibinfo {pages} {468} (\bibinfo {year} {2018})},\ \Eprint
  {http://arxiv.org/abs/1708.01921} {arXiv:1708.01921 [hep-th]} \BibitemShut
  {NoStop}%
\bibitem [{\citenamefont {Strickland}\ \emph {et~al.}(2017)\citenamefont
  {Strickland}, \citenamefont {Noronha},\ and\ \citenamefont
  {Denicol}}]{Strickland:2017kux}%
  \BibitemOpen
  \bibfield  {author} {\bibinfo {author} {\bibfnamefont {M.}~\bibnamefont
  {Strickland}}, \bibinfo {author} {\bibfnamefont {J.}~\bibnamefont {Noronha}},
  \ and\ \bibinfo {author} {\bibfnamefont {G.}~\bibnamefont {Denicol}},\
  }\href@noop {} {\  (\bibinfo {year} {2017})},\ \Eprint
  {http://arxiv.org/abs/1709.06644} {arXiv:1709.06644 [nucl-th]} \BibitemShut
  {NoStop}%
\bibitem [{\citenamefont
  {Romatschke}(2017{\natexlab{b}})}]{Romatschke:2017acs}%
  \BibitemOpen
  \bibfield  {author} {\bibinfo {author} {\bibfnamefont {P.}~\bibnamefont
  {Romatschke}},\ }\href {\doibase 10.1007/JHEP12(2017)079} {\bibfield
  {journal} {\bibinfo  {journal} {JHEP}\ }\textbf {\bibinfo {volume} {12}},\
  \bibinfo {pages} {079} (\bibinfo {year} {2017}{\natexlab{b}})},\ \Eprint
  {http://arxiv.org/abs/1710.03234} {arXiv:1710.03234 [hep-th]} \BibitemShut
  {NoStop}%
\bibitem [{\citenamefont {Denicol}\ and\ \citenamefont
  {Noronha}(2017)}]{Denicol:2017lxn}%
  \BibitemOpen
  \bibfield  {author} {\bibinfo {author} {\bibfnamefont {G.~S.}\ \bibnamefont
  {Denicol}}\ and\ \bibinfo {author} {\bibfnamefont {J.}~\bibnamefont
  {Noronha}},\ }\href@noop {} {\  (\bibinfo {year} {2017})},\ \Eprint
  {http://arxiv.org/abs/1711.01657} {arXiv:1711.01657 [nucl-th]} \BibitemShut
  {NoStop}%
\bibitem [{\citenamefont {Behtash}\ \emph {et~al.}(2017)\citenamefont
  {Behtash}, \citenamefont {Cruz-Camacho},\ and\ \citenamefont
  {Martinez}}]{Behtash:2017wqg}%
  \BibitemOpen
  \bibfield  {author} {\bibinfo {author} {\bibfnamefont {A.}~\bibnamefont
  {Behtash}}, \bibinfo {author} {\bibfnamefont {C.~N.}\ \bibnamefont
  {Cruz-Camacho}}, \ and\ \bibinfo {author} {\bibfnamefont {M.}~\bibnamefont
  {Martinez}},\ }\href@noop {} {\  (\bibinfo {year} {2017})},\ \Eprint
  {http://arxiv.org/abs/1711.01745} {arXiv:1711.01745 [hep-th]} \BibitemShut
  {NoStop}%
\bibitem [{\citenamefont {Grozdanov}\ \emph {et~al.}(2016)\citenamefont
  {Grozdanov}, \citenamefont {Kaplis},\ and\ \citenamefont
  {Starinets}}]{Grozdanov:2016vgg}%
  \BibitemOpen
  \bibfield  {author} {\bibinfo {author} {\bibfnamefont {S.}~\bibnamefont
  {Grozdanov}}, \bibinfo {author} {\bibfnamefont {N.}~\bibnamefont {Kaplis}}, \
  and\ \bibinfo {author} {\bibfnamefont {A.~O.}\ \bibnamefont {Starinets}},\
  }\href {\doibase 10.1007/JHEP07(2016)151} {\bibfield  {journal} {\bibinfo
  {journal} {JHEP}\ }\textbf {\bibinfo {volume} {07}},\ \bibinfo {pages} {151}
  (\bibinfo {year} {2016})},\ \Eprint {http://arxiv.org/abs/1605.02173}
  {arXiv:1605.02173 [hep-th]} \BibitemShut {NoStop}%
\bibitem [{\citenamefont {Andrade}\ \emph {et~al.}(2017)\citenamefont
  {Andrade}, \citenamefont {Casalderrey-Solana},\ and\ \citenamefont
  {Ficnar}}]{Andrade:2016rln}%
  \BibitemOpen
  \bibfield  {author} {\bibinfo {author} {\bibfnamefont {T.}~\bibnamefont
  {Andrade}}, \bibinfo {author} {\bibfnamefont {J.}~\bibnamefont
  {Casalderrey-Solana}}, \ and\ \bibinfo {author} {\bibfnamefont
  {A.}~\bibnamefont {Ficnar}},\ }\href {\doibase 10.1007/JHEP02(2017)016}
  {\bibfield  {journal} {\bibinfo  {journal} {JHEP}\ }\textbf {\bibinfo
  {volume} {02}},\ \bibinfo {pages} {016} (\bibinfo {year} {2017})},\ \Eprint
  {http://arxiv.org/abs/1610.08987} {arXiv:1610.08987 [hep-th]} \BibitemShut
  {NoStop}%
\bibitem [{\citenamefont {Grozdanov}\ and\ \citenamefont {van~der
  Schee}(2017)}]{Grozdanov:2016zjj}%
  \BibitemOpen
  \bibfield  {author} {\bibinfo {author} {\bibfnamefont {S.}~\bibnamefont
  {Grozdanov}}\ and\ \bibinfo {author} {\bibfnamefont {W.}~\bibnamefont
  {van~der Schee}},\ }\href {\doibase 10.1103/PhysRevLett.119.011601}
  {\bibfield  {journal} {\bibinfo  {journal} {Phys. Rev. Lett.}\ }\textbf
  {\bibinfo {volume} {119}},\ \bibinfo {pages} {011601} (\bibinfo {year}
  {2017})},\ \Eprint {http://arxiv.org/abs/1610.08976} {arXiv:1610.08976
  [hep-th]} \BibitemShut {NoStop}%
\bibitem [{\citenamefont {Grozdanov}\ and\ \citenamefont
  {Starinets}(2017)}]{Grozdanov:2016fkt}%
  \BibitemOpen
  \bibfield  {author} {\bibinfo {author} {\bibfnamefont {S.}~\bibnamefont
  {Grozdanov}}\ and\ \bibinfo {author} {\bibfnamefont {A.~O.}\ \bibnamefont
  {Starinets}},\ }\href {\doibase 10.1007/JHEP03(2017)166} {\bibfield
  {journal} {\bibinfo  {journal} {JHEP}\ }\textbf {\bibinfo {volume} {03}},\
  \bibinfo {pages} {166} (\bibinfo {year} {2017})},\ \Eprint
  {http://arxiv.org/abs/1611.07053} {arXiv:1611.07053 [hep-th]} \BibitemShut
  {NoStop}%
\bibitem [{\citenamefont {Casalderrey-Solana}\ \emph
  {et~al.}(2017)\citenamefont {Casalderrey-Solana}, \citenamefont {Gushterov},\
  and\ \citenamefont {Meiring}}]{Casalderrey-Solana:2017zyh}%
  \BibitemOpen
  \bibfield  {author} {\bibinfo {author} {\bibfnamefont {J.}~\bibnamefont
  {Casalderrey-Solana}}, \bibinfo {author} {\bibfnamefont {N.~I.}\ \bibnamefont
  {Gushterov}}, \ and\ \bibinfo {author} {\bibfnamefont {B.}~\bibnamefont
  {Meiring}},\ }\href@noop {} {\  (\bibinfo {year} {2017})},\ \Eprint
  {http://arxiv.org/abs/1712.02772} {arXiv:1712.02772 [hep-th]} \BibitemShut
  {NoStop}%
\bibitem [{\citenamefont {Keegan}\ \emph {et~al.}(2015)\citenamefont {Keegan},
  \citenamefont {Kurkela}, \citenamefont {Romatschke}, \citenamefont {van~der
  Schee},\ and\ \citenamefont {Zhu}}]{Keegan:2015avk}%
  \BibitemOpen
  \bibfield  {author} {\bibinfo {author} {\bibfnamefont {L.}~\bibnamefont
  {Keegan}}, \bibinfo {author} {\bibfnamefont {A.}~\bibnamefont {Kurkela}},
  \bibinfo {author} {\bibfnamefont {P.}~\bibnamefont {Romatschke}}, \bibinfo
  {author} {\bibfnamefont {W.}~\bibnamefont {van~der Schee}}, \ and\ \bibinfo
  {author} {\bibfnamefont {Y.}~\bibnamefont {Zhu}},\ }\href@noop {} {\
  (\bibinfo {year} {2015})},\ \Eprint {http://arxiv.org/abs/1512.05347}
  {arXiv:1512.05347 [hep-th]} \BibitemShut {NoStop}%
\bibitem [{\citenamefont {Keegan}\ \emph {et~al.}(2016)\citenamefont {Keegan},
  \citenamefont {Kurkela}, \citenamefont {Mazeliauskas},\ and\ \citenamefont
  {Teaney}}]{Keegan:2016cpi}%
  \BibitemOpen
  \bibfield  {author} {\bibinfo {author} {\bibfnamefont {L.}~\bibnamefont
  {Keegan}}, \bibinfo {author} {\bibfnamefont {A.}~\bibnamefont {Kurkela}},
  \bibinfo {author} {\bibfnamefont {A.}~\bibnamefont {Mazeliauskas}}, \ and\
  \bibinfo {author} {\bibfnamefont {D.}~\bibnamefont {Teaney}},\ }\href
  {\doibase 10.1007/JHEP08(2016)171} {\bibfield  {journal} {\bibinfo  {journal}
  {JHEP}\ }\textbf {\bibinfo {volume} {08}},\ \bibinfo {pages} {171} (\bibinfo
  {year} {2016})},\ \Eprint {http://arxiv.org/abs/1605.04287} {arXiv:1605.04287
  [hep-ph]} \BibitemShut {NoStop}%
\bibitem [{\citenamefont {Romatschke}\ and\ \citenamefont
  {Romatschke}(2017)}]{Romatschke:2017ejr}%
  \BibitemOpen
  \bibfield  {author} {\bibinfo {author} {\bibfnamefont {P.}~\bibnamefont
  {Romatschke}}\ and\ \bibinfo {author} {\bibfnamefont {U.}~\bibnamefont
  {Romatschke}},\ }\href@noop {} {\  (\bibinfo {year} {2017})},\ \Eprint
  {http://arxiv.org/abs/1712.05815} {arXiv:1712.05815 [nucl-th]} \BibitemShut
  {NoStop}%
\bibitem [{\citenamefont {Alqahtani}\ \emph {et~al.}(2017)\citenamefont
  {Alqahtani}, \citenamefont {Nopoush},\ and\ \citenamefont
  {Strickland}}]{Alqahtani:2017mhy}%
  \BibitemOpen
  \bibfield  {author} {\bibinfo {author} {\bibfnamefont {M.}~\bibnamefont
  {Alqahtani}}, \bibinfo {author} {\bibfnamefont {M.}~\bibnamefont {Nopoush}},
  \ and\ \bibinfo {author} {\bibfnamefont {M.}~\bibnamefont {Strickland}},\
  }\href@noop {} {\  (\bibinfo {year} {2017})},\ \Eprint
  {http://arxiv.org/abs/1712.03282} {arXiv:1712.03282 [nucl-th]} \BibitemShut
  {NoStop}%
\bibitem [{\citenamefont {Baym}(1984)}]{Baym:1984np}%
  \BibitemOpen
  \bibfield  {author} {\bibinfo {author} {\bibfnamefont {G.}~\bibnamefont
  {Baym}},\ }\href {\doibase 10.1016/0370-2693(84)91863-X} {\bibfield
  {journal} {\bibinfo  {journal} {Phys. Lett.}\ }\textbf {\bibinfo {volume}
  {B138}},\ \bibinfo {pages} {18} (\bibinfo {year} {1984})}\BibitemShut
  {NoStop}%
\bibitem [{\citenamefont {Florkowski}\ \emph {et~al.}(2013)\citenamefont
  {Florkowski}, \citenamefont {Ryblewski},\ and\ \citenamefont
  {Strickland}}]{Florkowski:2013lya}%
  \BibitemOpen
  \bibfield  {author} {\bibinfo {author} {\bibfnamefont {W.}~\bibnamefont
  {Florkowski}}, \bibinfo {author} {\bibfnamefont {R.}~\bibnamefont
  {Ryblewski}}, \ and\ \bibinfo {author} {\bibfnamefont {M.}~\bibnamefont
  {Strickland}},\ }\href {\doibase 10.1103/PhysRevC.88.024903} {\bibfield
  {journal} {\bibinfo  {journal} {Phys. Rev.}\ }\textbf {\bibinfo {volume}
  {C88}},\ \bibinfo {pages} {024903} (\bibinfo {year} {2013})},\ \Eprint
  {http://arxiv.org/abs/1305.7234} {arXiv:1305.7234 [nucl-th]} \BibitemShut
  {NoStop}%
\bibitem [{\citenamefont {Bhatnagar}\ \emph {et~al.}(1954)\citenamefont
  {Bhatnagar}, \citenamefont {Gross},\ and\ \citenamefont
  {Krook}}]{Bhatnagar:1954zz}%
  \BibitemOpen
  \bibfield  {author} {\bibinfo {author} {\bibfnamefont {P.~L.}\ \bibnamefont
  {Bhatnagar}}, \bibinfo {author} {\bibfnamefont {E.~P.}\ \bibnamefont
  {Gross}}, \ and\ \bibinfo {author} {\bibfnamefont {M.}~\bibnamefont
  {Krook}},\ }\href {\doibase 10.1103/PhysRev.94.511} {\bibfield  {journal}
  {\bibinfo  {journal} {Phys. Rev.}\ }\textbf {\bibinfo {volume} {94}},\
  \bibinfo {pages} {511} (\bibinfo {year} {1954})}\BibitemShut {NoStop}%
\bibitem [{\citenamefont {Anderson}\ and\ \citenamefont
  {Witting}(1974)}]{Anderson:1974a}%
  \BibitemOpen
  \bibfield  {author} {\bibinfo {author} {\bibfnamefont {J.~L.}\ \bibnamefont
  {Anderson}}\ and\ \bibinfo {author} {\bibfnamefont {H.~R.}\ \bibnamefont
  {Witting}},\ }\href@noop {} {\bibfield  {journal} {\bibinfo  {journal}
  {Physica}\ }\textbf {\bibinfo {volume} {74}},\ \bibinfo {pages} {466}
  (\bibinfo {year} {1974})}\BibitemShut {NoStop}%
\bibitem [{\citenamefont {Heller}\ \emph {et~al.}(2012)\citenamefont {Heller},
  \citenamefont {Janik},\ and\ \citenamefont {Witaszczyk}}]{Heller:2011ju}%
  \BibitemOpen
  \bibfield  {author} {\bibinfo {author} {\bibfnamefont {M.~P.}\ \bibnamefont
  {Heller}}, \bibinfo {author} {\bibfnamefont {R.~A.}\ \bibnamefont {Janik}}, \
  and\ \bibinfo {author} {\bibfnamefont {P.}~\bibnamefont {Witaszczyk}},\
  }\href {\doibase 10.1103/PhysRevLett.108.201602} {\bibfield  {journal}
  {\bibinfo  {journal} {Phys.Rev.Lett.}\ }\textbf {\bibinfo {volume} {108}},\
  \bibinfo {pages} {201602} (\bibinfo {year} {2012})},\ \Eprint
  {http://arxiv.org/abs/1103.3452} {arXiv:1103.3452 [hep-th]} \BibitemShut
  {NoStop}%
\bibitem [{\citenamefont {Denicol}\ \emph
  {et~al.}(2014{\natexlab{a}})\citenamefont {Denicol}, \citenamefont {Heinz},
  \citenamefont {Martinez}, \citenamefont {Noronha},\ and\ \citenamefont
  {Strickland}}]{Denicol:2014tha}%
  \BibitemOpen
  \bibfield  {author} {\bibinfo {author} {\bibfnamefont {G.~S.}\ \bibnamefont
  {Denicol}}, \bibinfo {author} {\bibfnamefont {U.~W.}\ \bibnamefont {Heinz}},
  \bibinfo {author} {\bibfnamefont {M.}~\bibnamefont {Martinez}}, \bibinfo
  {author} {\bibfnamefont {J.}~\bibnamefont {Noronha}}, \ and\ \bibinfo
  {author} {\bibfnamefont {M.}~\bibnamefont {Strickland}},\ }\href {\doibase
  10.1103/PhysRevD.90.125026} {\bibfield  {journal} {\bibinfo  {journal} {Phys.
  Rev.}\ }\textbf {\bibinfo {volume} {D90}},\ \bibinfo {pages} {125026}
  (\bibinfo {year} {2014}{\natexlab{a}})},\ \Eprint
  {http://arxiv.org/abs/1408.7048} {arXiv:1408.7048 [hep-ph]} \BibitemShut
  {NoStop}%
\bibitem [{\citenamefont {Denicol}\ \emph
  {et~al.}(2014{\natexlab{b}})\citenamefont {Denicol}, \citenamefont {Heinz},
  \citenamefont {Martinez}, \citenamefont {Noronha},\ and\ \citenamefont
  {Strickland}}]{Denicol:2014xca}%
  \BibitemOpen
  \bibfield  {author} {\bibinfo {author} {\bibfnamefont {G.~S.}\ \bibnamefont
  {Denicol}}, \bibinfo {author} {\bibfnamefont {U.~W.}\ \bibnamefont {Heinz}},
  \bibinfo {author} {\bibfnamefont {M.}~\bibnamefont {Martinez}}, \bibinfo
  {author} {\bibfnamefont {J.}~\bibnamefont {Noronha}}, \ and\ \bibinfo
  {author} {\bibfnamefont {M.}~\bibnamefont {Strickland}},\ }\href {\doibase
  10.1103/PhysRevLett.113.202301} {\bibfield  {journal} {\bibinfo  {journal}
  {Phys. Rev. Lett.}\ }\textbf {\bibinfo {volume} {113}},\ \bibinfo {pages}
  {202301} (\bibinfo {year} {2014}{\natexlab{b}})},\ \Eprint
  {http://arxiv.org/abs/1408.5646} {arXiv:1408.5646 [hep-ph]} \BibitemShut
  {NoStop}%
\bibitem [{\citenamefont {Romatschke}(2018)}]{Romatschke:2018wgi}%
  \BibitemOpen
  \bibfield  {author} {\bibinfo {author} {\bibfnamefont {P.}~\bibnamefont
  {Romatschke}},\ }\href@noop {} {\  (\bibinfo {year} {2018})},\ \Eprint
  {http://arxiv.org/abs/1802.06804} {arXiv:1802.06804 [nucl-th]} \BibitemShut
  {NoStop}%
\bibitem [{\citenamefont {Basar}\ \emph {et~al.}(2013)\citenamefont {Basar},
  \citenamefont {Dunne},\ and\ \citenamefont {Unsal}}]{Basar:2013eka}%
  \BibitemOpen
  \bibfield  {author} {\bibinfo {author} {\bibfnamefont {G.}~\bibnamefont
  {Basar}}, \bibinfo {author} {\bibfnamefont {G.~V.}\ \bibnamefont {Dunne}}, \
  and\ \bibinfo {author} {\bibfnamefont {M.}~\bibnamefont {Unsal}},\ }\href
  {\doibase 10.1007/JHEP10(2013)041} {\bibfield  {journal} {\bibinfo  {journal}
  {JHEP}\ }\textbf {\bibinfo {volume} {10}},\ \bibinfo {pages} {041} (\bibinfo
  {year} {2013})},\ \Eprint {http://arxiv.org/abs/1308.1108} {arXiv:1308.1108
  [hep-th]} \BibitemShut {NoStop}%
\bibitem [{\citenamefont {{Yamada}}\ and\ \citenamefont
  {{Ikeda}}(2013)}]{PadeCut}%
  \BibitemOpen
  \bibfield  {author} {\bibinfo {author} {\bibfnamefont {H.~S.}\ \bibnamefont
  {{Yamada}}}\ and\ \bibinfo {author} {\bibfnamefont {K.~S.}\ \bibnamefont
  {{Ikeda}}},\ }\href@noop {} {\  (\bibinfo {year} {2013})},\ \Eprint
  {http://arxiv.org/abs/1308.4453} {arXiv:1308.4453 [math-ph]} \BibitemShut
  {NoStop}%
\bibitem [{\citenamefont {Dunne}()}]{Dunne}%
  \BibitemOpen
  \bibfield  {author} {\bibinfo {author} {\bibfnamefont {G.~V.}\ \bibnamefont
  {Dunne}},\ }\href@noop {} {\bibinfo  {journal} {Lectures given at the
  Schladming Winter School 2015}\ }\BibitemShut {NoStop}%
\bibitem [{\citenamefont {Aniceto}\ \emph {et~al.}(2018)\citenamefont
  {Aniceto}, \citenamefont {Basar},\ and\ \citenamefont
  {Schiappa}}]{Aniceto:2018bis}%
  \BibitemOpen
\bibfield  {journal} {  }\bibfield  {author} {\bibinfo {author} {\bibfnamefont
  {I.}~\bibnamefont {Aniceto}}, \bibinfo {author} {\bibfnamefont
  {G.}~\bibnamefont {Basar}}, \ and\ \bibinfo {author} {\bibfnamefont
  {R.}~\bibnamefont {Schiappa}},\ }\href@noop {} {\  (\bibinfo {year}
  {2018})},\ \Eprint {http://arxiv.org/abs/1802.10441} {arXiv:1802.10441
  [hep-th]} \BibitemShut {NoStop}%
\bibitem [{\citenamefont {Aniceto}(2016)}]{Aniceto:2015rua}%
  \BibitemOpen
  \bibfield  {author} {\bibinfo {author} {\bibfnamefont {I.}~\bibnamefont
  {Aniceto}},\ }\href {\doibase 10.1088/1751-8113/49/6/065403} {\bibfield
  {journal} {\bibinfo  {journal} {J. Phys.}\ }\textbf {\bibinfo {volume}
  {A49}},\ \bibinfo {pages} {065403} (\bibinfo {year} {2016})},\ \Eprint
  {http://arxiv.org/abs/1506.03388} {arXiv:1506.03388 [hep-th]} \BibitemShut
  {NoStop}%
\bibitem [{\citenamefont {{Dutykh}}(2016)}]{Dutykh:2016spectral}%
  \BibitemOpen
  \bibfield  {author} {\bibinfo {author} {\bibfnamefont {D.}~\bibnamefont
  {{Dutykh}}},\ }\href@noop {} {\  (\bibinfo {year} {2016})},\ \Eprint
  {http://arxiv.org/abs/1606.05432} {arXiv:1606.05432 [math.NA]} \BibitemShut
  {NoStop}%
\bibitem [{\citenamefont {Boyd}(2001)}]{boyd2001chebyshev}%
  \BibitemOpen
  \bibfield  {author} {\bibinfo {author} {\bibfnamefont {J.~P.}\ \bibnamefont
  {Boyd}},\ }\href@noop {} {\emph {\bibinfo {title} {Chebyshev and Fourier
  spectral methods}}}\ (\bibinfo  {publisher} {Courier Corporation},\ \bibinfo
  {year} {2001})\BibitemShut {NoStop}%
\bibitem [{\citenamefont {Liu}\ \emph {et~al.}(2015)\citenamefont {Liu},
  \citenamefont {Shen},\ and\ \citenamefont {Heinz}}]{Liu:2015nwa}%
  \BibitemOpen
  \bibfield  {author} {\bibinfo {author} {\bibfnamefont {J.}~\bibnamefont
  {Liu}}, \bibinfo {author} {\bibfnamefont {C.}~\bibnamefont {Shen}}, \ and\
  \bibinfo {author} {\bibfnamefont {U.}~\bibnamefont {Heinz}},\ }\href
  {\doibase 10.1103/PhysRevC.92.049904, 10.1103/PhysRevC.91.064906} {\bibfield
  {journal} {\bibinfo  {journal} {Phys. Rev.}\ }\textbf {\bibinfo {volume}
  {C91}},\ \bibinfo {pages} {064906} (\bibinfo {year} {2015})},\ \bibinfo
  {note} {[Erratum: Phys. Rev.C92,no.4,049904(2015)]},\ \Eprint
  {http://arxiv.org/abs/1504.02160} {arXiv:1504.02160 [nucl-th]} \BibitemShut
  {NoStop}%
\bibitem [{\citenamefont {Brewer}\ and\ \citenamefont
  {Romatschke}(2015)}]{Brewer:2015ipa}%
  \BibitemOpen
  \bibfield  {author} {\bibinfo {author} {\bibfnamefont {J.}~\bibnamefont
  {Brewer}}\ and\ \bibinfo {author} {\bibfnamefont {P.}~\bibnamefont
  {Romatschke}},\ }\href {\doibase 10.1103/PhysRevLett.115.190404} {\bibfield
  {journal} {\bibinfo  {journal} {Phys. Rev. Lett.}\ }\textbf {\bibinfo
  {volume} {115}},\ \bibinfo {pages} {190404} (\bibinfo {year} {2015})},\
  \Eprint {http://arxiv.org/abs/1508.01199} {arXiv:1508.01199 [hep-th]}
  \BibitemShut {NoStop}%
\bibitem [{\citenamefont {Bazow}\ \emph {et~al.}(2016)\citenamefont {Bazow},
  \citenamefont {Denicol}, \citenamefont {Heinz}, \citenamefont {Martinez},\
  and\ \citenamefont {Noronha}}]{Bazow:2015dha}%
  \BibitemOpen
  \bibfield  {author} {\bibinfo {author} {\bibfnamefont {D.}~\bibnamefont
  {Bazow}}, \bibinfo {author} {\bibfnamefont {G.~S.}\ \bibnamefont {Denicol}},
  \bibinfo {author} {\bibfnamefont {U.}~\bibnamefont {Heinz}}, \bibinfo
  {author} {\bibfnamefont {M.}~\bibnamefont {Martinez}}, \ and\ \bibinfo
  {author} {\bibfnamefont {J.}~\bibnamefont {Noronha}},\ }\href {\doibase
  10.1103/PhysRevLett.116.022301} {\bibfield  {journal} {\bibinfo  {journal}
  {Phys. Rev. Lett.}\ }\textbf {\bibinfo {volume} {116}},\ \bibinfo {pages}
  {022301} (\bibinfo {year} {2016})},\ \Eprint
  {http://arxiv.org/abs/1507.07834} {arXiv:1507.07834 [hep-ph]} \BibitemShut
  {NoStop}%
\bibitem [{\citenamefont {Buchel}\ \emph {et~al.}(2016)\citenamefont {Buchel},
  \citenamefont {Heller},\ and\ \citenamefont {Noronha}}]{Buchel:2016cbj}%
  \BibitemOpen
  \bibfield  {author} {\bibinfo {author} {\bibfnamefont {A.}~\bibnamefont
  {Buchel}}, \bibinfo {author} {\bibfnamefont {M.~P.}\ \bibnamefont {Heller}},
  \ and\ \bibinfo {author} {\bibfnamefont {J.}~\bibnamefont {Noronha}},\ }\href
  {\doibase 10.1103/PhysRevD.94.106011} {\bibfield  {journal} {\bibinfo
  {journal} {Phys. Rev.}\ }\textbf {\bibinfo {volume} {D94}},\ \bibinfo {pages}
  {106011} (\bibinfo {year} {2016})},\ \Eprint
  {http://arxiv.org/abs/1603.05344} {arXiv:1603.05344 [hep-th]} \BibitemShut
  {NoStop}%
\bibitem [{\citenamefont {Gubser}(2010)}]{Gubser:2010ze}%
  \BibitemOpen
  \bibfield  {author} {\bibinfo {author} {\bibfnamefont {S.~S.}\ \bibnamefont
  {Gubser}},\ }\href {\doibase 10.1103/PhysRevD.82.085027} {\bibfield
  {journal} {\bibinfo  {journal} {Phys. Rev.}\ }\textbf {\bibinfo {volume}
  {D82}},\ \bibinfo {pages} {085027} (\bibinfo {year} {2010})},\ \Eprint
  {http://arxiv.org/abs/1006.0006} {arXiv:1006.0006 [hep-th]} \BibitemShut
  {NoStop}%
\end{thebibliography}%
\end{document}